\documentclass[10pt,twocolumn,aps,superscriptaddress,showpacs,floatfix,prd,noshowpacs]{revtex4-2}
\usepackage{mathrsfs}
\usepackage{amssymb}
\usepackage{amsmath}
\usepackage{graphicx}
\usepackage[normalem]{ulem}
\usepackage[dvips]{color}
\usepackage{bm}
\usepackage{longtable}
\usepackage{slashed}
\usepackage{enumitem}

\usepackage{titlesec}
\renewcommand{\thesection}{\Roman{section}}
\titleformat{\section}{\small\bfseries\centering}{\thesection.}{0.5em}{}

\usepackage{times}
\usepackage{fouriernc}

\usepackage[bb=ams, cal=cm, scr=boondox, frak=euler]{mathalpha}

\usepackage[breaklinks=true]{hyperref}
\hypersetup{
  colorlinks=true,
  citecolor=magenta,
  linkcolor=black,
  urlcolor=magenta
}

\renewcommand\sout{\bgroup \color{red} \ULdepth=-.5ex \ULset}

\renewcommand{\rm}[1]{\textrm{#1}}
\renewcommand{\d}{\mathrm{d}}

\begin{document}
\title{Unraveling Trace Anomaly of Supradense Matter via Neutron Star Compactness Scaling}

\author{Bao-Jun Cai\footnote{bjcai@fudan.edu.cn}}
\affiliation{Key Laboratory of Nuclear Physics and Ion-beam Application (MOE), Institute of Modern Physics, Fudan University, Shanghai 200433, China} 
\affiliation{Shanghai Research Center for Theoretical Nuclear Physics,
NSFC and Fudan University, Shanghai 200438, China}
\author{Bao-An Li\footnote{Bao-An.Li@etamu.edu}}
\affiliation{Department of Physics and Astronomy, East Texas A$\&$M
University, Commerce, TX 75429-3011, USA}

\date{\today}

\newcommand{\x}{\mathrm{X}}
\newcommand{\y}{\mathrm{Y}}

\begin{abstract}
The trace anomaly $\Delta\equiv 1/3-P/\varepsilon=1/3-\phi$ quantifies the possibly broken conformal symmetry in supradense matter under pressure $P$ at energy density $\varepsilon$. 
Perturbative QCD (pQCD) predicts a vanishing $\Delta$ at extremely high energy or baryon densities when the conformal symmetry is realized but its behavior at intermediate densities reachable in neutron stars (NSs) are still very uncertain. The extraction of $\Delta$ from NS observations strongly depends on the employed model for nuclear Equation of State (EOS). Using the IPAD-TOV method based on an Intrinsic and Perturbatively Analysis of the Dimensionless (IPAD) Tolman-Oppenheimer-Volkoff (TOV) equations that are further verified numerically by using $10^5$ EOSs generated randomly with a meta-model in a very broad EOS parameter space constrained by terrestrial nuclear experiments and astrophysical observations, 
here we first show that the compactness $\xi\equiv GM_{\rm{NS}}/Rc^2\equiv M_{\rm{NS}}/R$ of a NS with mass $M_{\rm{NS}}$ and radius $R$ scales very accurately with $\overline{\Pi}_{\rm{c}}\equiv\Pi_{\rm{c}}\cdot(1+18\x/25)\equiv\x/(1+3\x^2+4\x)\cdot(1+18\rm{X}/25)$ where $\x\equiv\phi_{\rm{c}}= P_{\rm{c}}/\varepsilon_{\rm{c}}$ is the ratio of pressure over energy density at NS centers.  The scaling of NS compactness thus enables one to readily read off the central trace anomaly $\Delta_{\rm{c}}=1/3-\x$ directly from the observational data of either the mass-radius or red-shift measurements. We then demonstrate indeed that the available NS data themselves from recent X-ray and gravitational wave observations can determine model-insensitively the trace anomaly as a function of energy density in NS cores, providing a stringent test of existing NS models and a clear guidance in a new direction for further understanding the nature and EOS of supradense matter.
\end{abstract}

\maketitle

\section{Introduction}

To understand the nature and Equation of State (EOS) of supradense matter existing in neutron stars (NSs) has been an important and long-standing scientific goal shared by both nuclear physics and astrophysics\,\cite{Walecka1974,Chin1976,Freedman1977,Akmal1998,LP01,Alford2008,LCK08}. 
The EOS at zero temperature is defined as the functional relationship $P(\varepsilon)$ between the pressure $P$ and energy density $\varepsilon$.
Thanks to extensive investigations\,\cite{Tews18,Baym19,McL19,Zhao20,Tan22,Tan22-a,Alt22,Dri22,Huang22,Ecker22,Ecker22-a,Musolino2024,Pro23,Som23,
Ann18,Ann23,ZLi23,Cao23,Mro23,Ess21,Brandes23,Brandes23-a,Tak23,Pang23,Fan23,Bed15,Olii23} utilizing various experimental and observational data especially since GW170817, much progress has been made, see Refs.\,\cite{Wat16,Oertel2017,Baym18,Isa18,LCXZ21,Dri21,Lovato22,Soren2023} for recent reviews. However, many interesting issues remain to be settled mostly because of the model dependences and degeneracies in analyzing various observables. In particular, characterizing the EOS and reflecting the nature of supradense matter, the trace anomaly $\Delta\equiv 1/3-P/\varepsilon=1/3-\phi$ measures the degree of conformal symmetry. The latter is expected to be fully realized with $\Delta=0$ at extremely high densities according to perturbative QCD (pQCD)\,\cite{Fuji22}. NSs are natural laboratories for testing such predictions about supradense matter.  Unfortunately, the information extracted so far about the trace anomaly from analyzing NS observables are still rather EOS model dependent.

Is there an essentially model-insensitive way enabling us to extract reliably the $\Delta$ solely from the NS observational data? Yes, in this work we show that the accurate scaling of NS compactness $\xi\equiv GM_{\rm{NS}}/Rc^2\equiv  M_{\rm{NS}}/R$ ($c=G=1$) with its central pressure/energy density ratio $\x\equiv P_{\rm{c}}/\varepsilon_{\rm{c}}$ allows us to do so.  This scaling has the advantage of largely canceling out uncertainties involved in both the mass and radius scalings. We find that one can easily read off the central trace anomaly $\Delta_{\rm{c}}$ directly from the observed compactness $\xi$. In particular, we demonstrate that the joint mass-radius observations for PSR J0030+0451\,\cite{Riley19,Miller19},  PSR J0740+6620\,\cite{Fon21,Riley21,Miller21,Salmi22} and PRS J0437-4715\,\cite{Choud24,Reardon24} by NICER (Neutron Star Interior Composition
Explorer),  the surface gravitational red-shift measurement of the NS in X-ray burster GS 1826-24\,\cite{Zhou23}, the mass-radius constraints for the two NSs involved in GW 170817\,\cite{Abbott2017,Abbott2018} and GW 190425\,\cite{Abbott2020-a},  respectively, and the redback spider pulsar PSR J2215+5135 with a mass about $2.15_{-0.10}^{+0.10}M_{\odot}$ ($M_{\odot}$=solar mass)\,\cite{Sul24} via a joint X-ray and optical analysis together determine model-insensitively the NS central trace anomaly as a function of energy density, enabling a stringent test of existing EOS models and pointing out a new direction for further investigating the trace anomaly of supradense matter.

The rest of this paper is organized as follows: Section \ref{sec_scalings} presents the scaling relations for NS mass and compactness, derived through the IPAD-TOV method\,\cite{CL25} by performing an Intrinsic and Perturbative Analysis of the Dimensionless (IPAD) Tolman-Oppenheimer-Volkoff (TOV) equations for reduced NS variables\,\cite{TOV39-1,TOV39-2}. Notably, we discuss effects on the compactness of a high-order correction term of ``$18\x/25$'' to the mass scaling. In Section \ref{sec_ta}, these scalings are applied to extract the central dimensionless trace anomaly and central energy density for several NS instances in a model-insensitive manner. Section \ref{sec_s2} explores whether current NS observations consistently support a peaked speed of sound profile in NSs. Section \ref{sec_conclusion} provides a summary of the present work and some perspectives of future studies using the approach established here. Three appendices are used to provide more technical details: Appendix \ref{app-meta} describes the meta EOS model, Appendix \ref{app-correction} details the derivation of the ``$18\x/25$'' correction term, and Appendix \ref{app-s2} presents analytical analyses of the speed of sound profile in NSs based on the trace anomaly decomposition.

\section{Scalings of NS Compactness and Mass from IPAD-TOV and Their Verification Using $10^5$ Meta-model EOSs}\label{sec_scalings}

Using the IPAD-TOV approach\,\cite{CLZ23-a,CLZ23-b,CL24,CL25}, we obtained previously the scalings of NS mass $M_{\rm{NS}}$ and radius $R$ as $M_{\rm{NS}}\propto\Gamma_{\rm{c}}\equiv\varepsilon_{\rm{c}}^{1/2}\Pi_{\rm{c}}^{3/2}$ and $R\propto\nu_{\rm{c}}\equiv \varepsilon_{\rm{c}}^{1/2}\Pi_{\rm{c}}^{1/2}$, where:
\begin{equation}\label{def_Pic}
\Pi_{\rm{c}}=\frac{\x}{1+3\x^2+4\x},
\end{equation}
here $\Gamma_{\rm{c}}$ and $\nu_{\rm{c}}$ are measured in $\rm{fm}^{3/2}/\rm{MeV}^{1/2}$ and $\Pi_{\rm{c}}$ is dimensionless.
Applying the scalings $M_{\rm{NS}}\propto\Gamma_{\rm{c}}$ and $R\propto\nu_{\rm{c}}$ to the TOV configuration at $M_{\rm{TOV}}\equiv M_{\rm{NS}}^{\max}$ using scaling coefficients determined by solving the original TOV equations with 104 most widely used NS EOSs (both microscopic and phenomenological) leads to a model-insensitive constraint on the EOS of the densest matter existing in our Universe\,\cite{CLZ23-a}. In Appendix \ref{app-correction}, we shall further verify quantitatively the compactness scaling at M$_{\rm{TOV}}$ by using totally 284 realistic EOSs including additionally 180 EOSs with more diverse features from the literature. Briefly addressed, the full ensemble now includes EOSs encapsulating a first-order phase transition, a hadron-quark crossover, several hyperons and/or $\Delta$ resonances, as well as those incorporating phenomenologically the speed of sound squared $s^2$ having multiple peaks or discontinuities due to more exotic physics predicted. It is interesting to note that very recently the above mass and radius scalings were independently verified by Lattimer quantitatively by using several hundred NS EOSs available in the literature\,\cite{Lat24-talk}. It was found that at $M_{\rm{TOV}}$, accuracies of the above mass and radius scalings are 7\% and 8\%; and at $1.4M_{\odot}$, they are 2\% and 6\%, respectively, with respect to solutions of the original TOV equations using the traditional approach\,\cite{Lat24-talk}. 
Similar scalings of $M_{\rm{NS}}$ and $R$ for NSs at the TOV configuration were recently studied also in Refs.\,\cite{Ofeng20,Ofeng24,SL24} from different starting points. All these studies offer useful and generally consistent approaches using mass and radius scalings for extracting directly the central EOS of NSs from their observational data.

\renewcommand*\figurename{\small FIG.}
\begin{figure*}
\centering
\includegraphics[height=6.8cm]{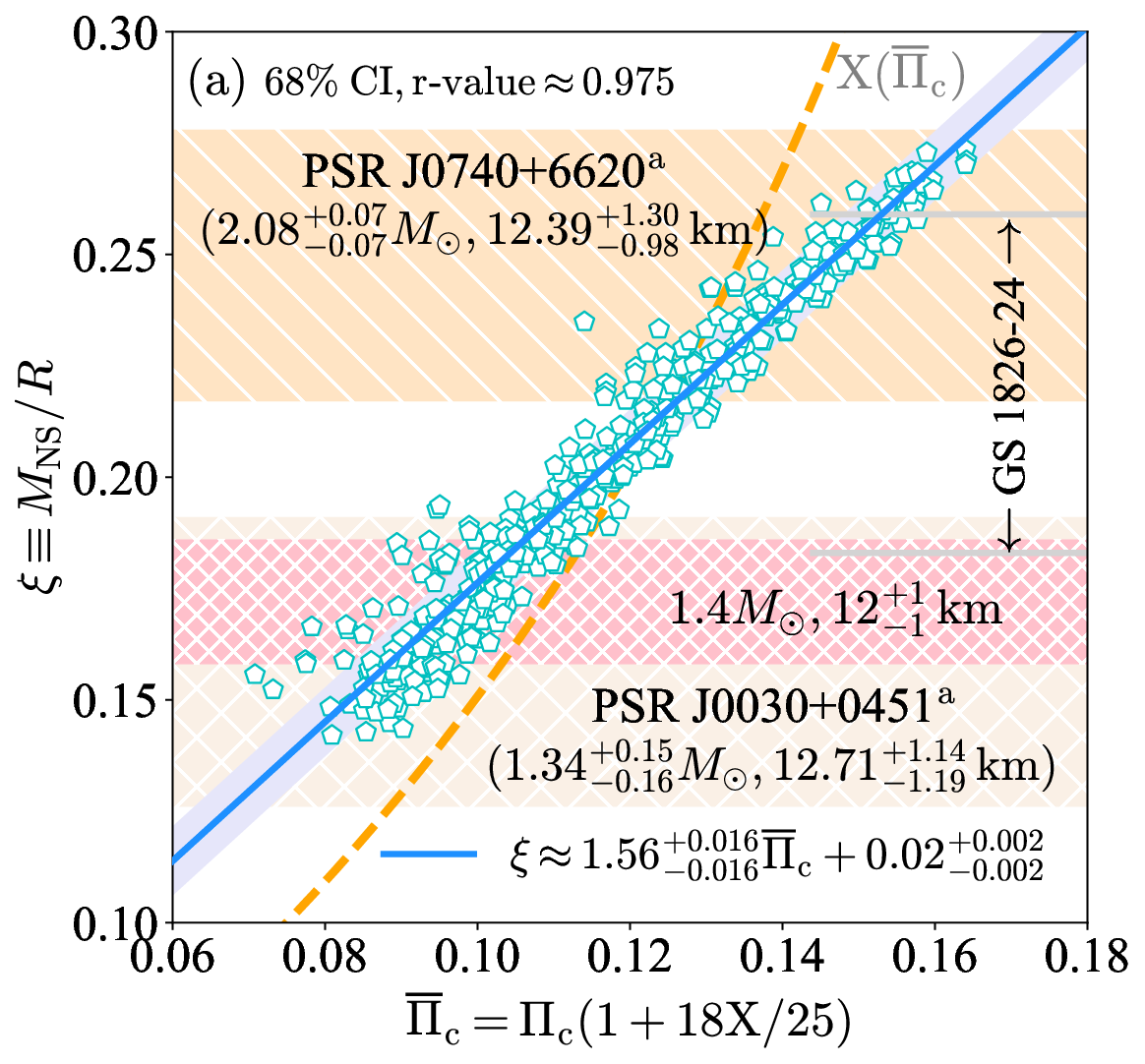}\qquad
\includegraphics[height=6.8cm]{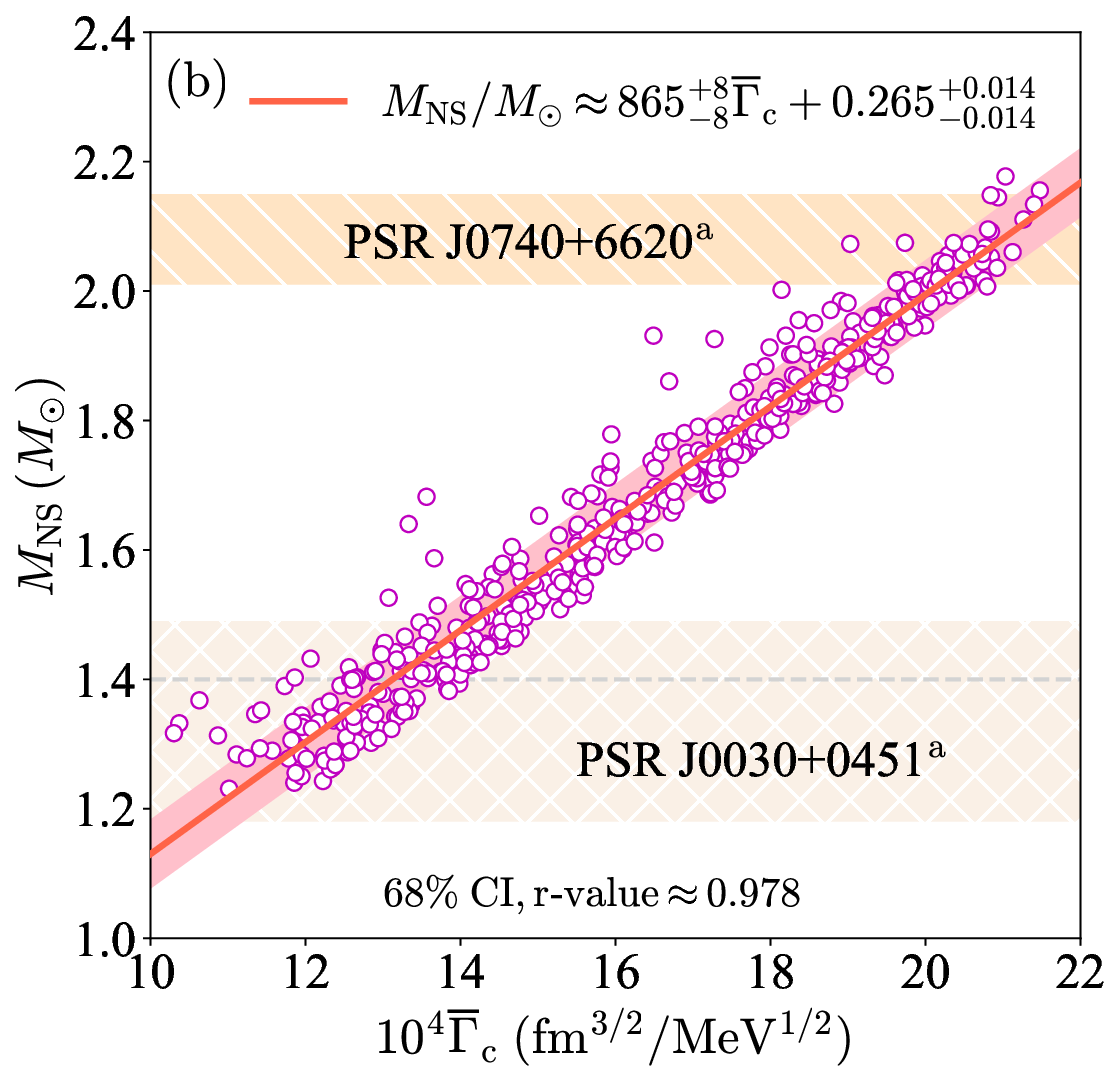}\\[0.25cm]
\includegraphics[height=6.8cm]{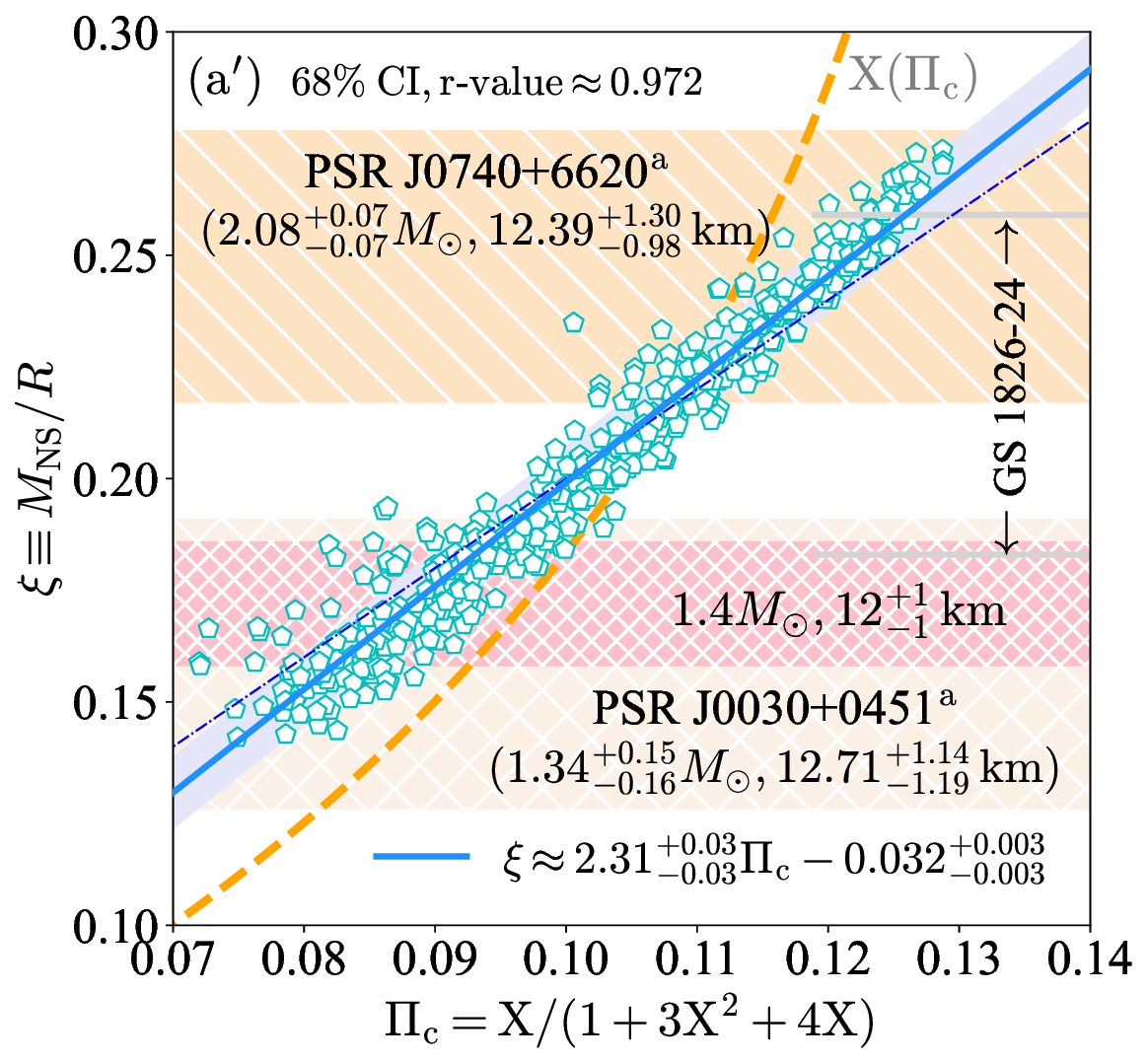}\qquad
\includegraphics[height=6.8cm]{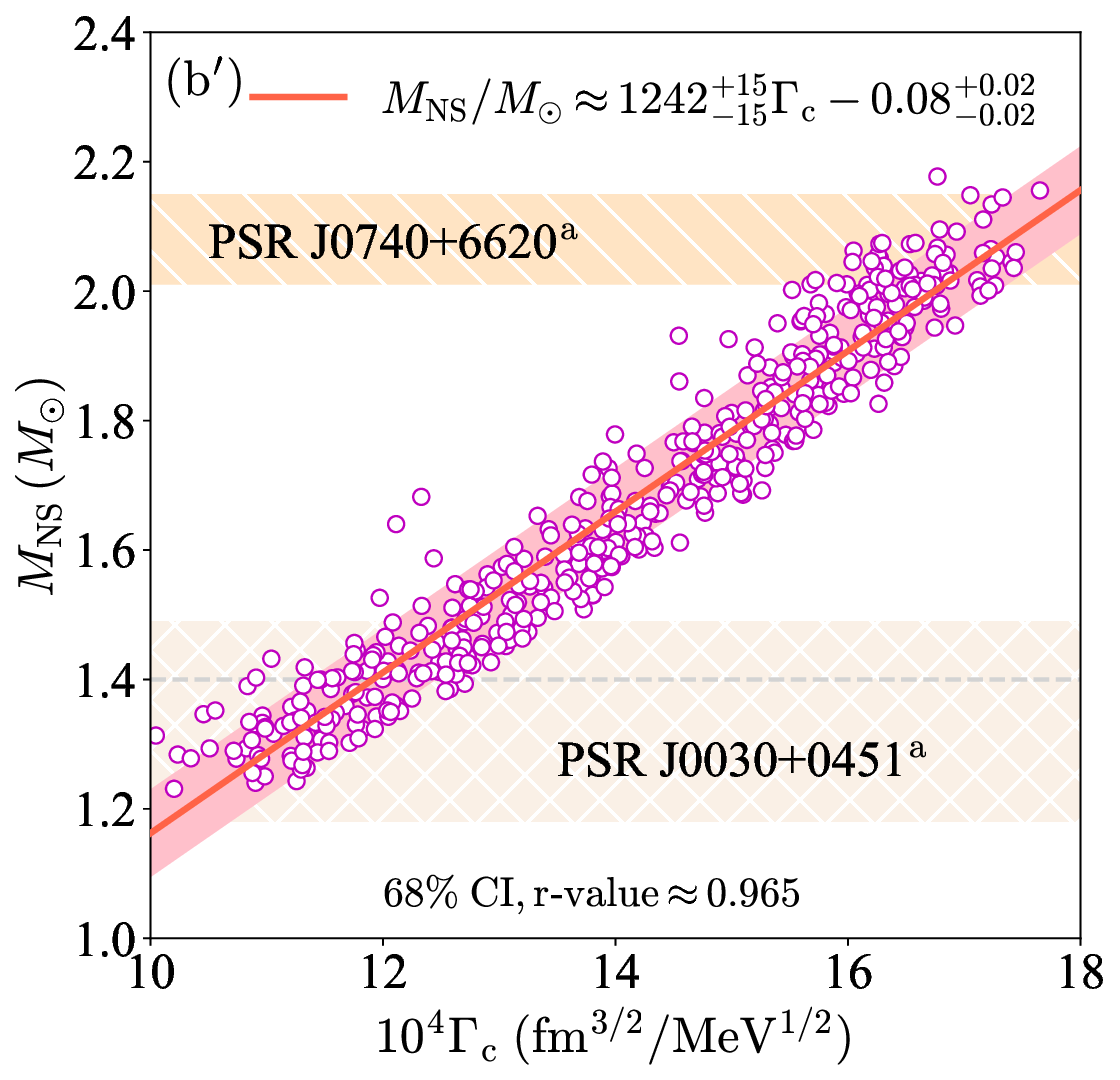}
\caption{(Color Online). Upper left panel: The revised compactness scaling $\xi$-$\overline{\Pi}_{\rm{c}}$ using meta-model EOSs consistent with observational/experimental constraints.  The compactnesses for PSR J0740+6620\,\cite{Riley21}, PSR J0030+0451\,\cite{Riley19} and the NS in the X-ray burster GS 1826-24\,\cite{Zhou23} are shown individually. The function $\x(\overline{\Pi}_{\rm{c}})$ is plotted by the dashed orange line and the compactness for a typical canonical NS\,\cite{Brandes23-a,Rich23} (with $R\approx12_{-1}^{+1}\,\rm{km}$) by the hatched pink band.  
Upper right panel: Same as panel (a) but for the revised mass scaling $M_{\rm{NS}}$-$\overline{\Gamma}_{\rm{c}}$, the lavender and pink bands in panel (a) and (b) represents their 68\% confidence intervals (CIs).
Lower two panels: The original compactness $\xi$-$\Pi_{\rm{c}}$ and mass $M_{\rm{NS}}$-$\Gamma_{\rm{c}}$ scalings.
}\label{fig_Pic}
\end{figure*}

The radius scaling was previously derived by truncating the expansion of reduced NS pressure at the second order of the reduced distance $\widehat{r}$ from the NS center (see Appendix \ref{app-correction}); while the NS mass was obtained effectively from the relation $M_{\rm{NS}}\sim\mbox{``central energy density''}\times R^3$\,\cite{CLZ23-a,CLZ23-b,CL24} which is the leading term in expanding the NS mass as a function of the reduced radius. Expanding the mass to the same order as the pressure, the NS mass scaling is then revised to
\begin{align}
M_{\rm{NS}}\propto&\overline{\Gamma}_{\rm{c}}=\Gamma_{\rm{c}}\left(1+\frac{18}{25}\x\right).\label{sca-M}
\end{align}
The high-order correction ``$18\rm{X}/25$'' to the original mass scaling is directly from analyzing the scaled TOV equations themselves instead of any fitting procedure;
we derive this correction term in detail in Appendix \ref{app-correction}.
The scaling for $R$ remains the same as $R\propto\nu_{\rm{c}}$. 
Consequently, the revised NS compactness scaling becomes
\begin{equation}
\xi\equiv\frac{M_{\rm{NS}}}{R}\propto
\overline{\Pi}_{\rm{c}}=\Pi_{\rm{c}}\left(1+\frac{18}{25}\x\right).\label{sca-xi}
\end{equation}
As shown in Appendix \ref{app-correction},  the relation (\ref{sca-xi}) significantly improves the compactness scaling for NSs at the TOV configuration using the aforementioned 284 realistic EOSs.

We notice that the TOV configuration predicted for a given EOS corresponds to a special state on the mass-radius sequence (e.g., the densest visible matter in the Universe predicted with the given EOS). However, it is probably just a hypothetical state that is not reachable for most NSs observed. In particular, any realistic NS EOS should generate a $M_{\rm{TOV}}$ at least greater than the mass of the currently observed most massive NSs, e.g., $M_{\rm{TOV}}\gtrsim 2.15M_{\odot}$ considering PSR J2215+5135\,\cite{Sul24}.
In this sense, a canonical NS could not be generally considered as at the TOV configuration. In order to obtain properties of NSs with $M_{\rm{NS}}\lesssim2M_{\odot}$ (including the canonical ones) using their observed masses/radii, we may necessarily rely on the scalings holding for generally stable NSs instead of the ones at the TOV configuration. In fact, one expects naturally that the scalings for these two classes of NSs (TOV NSs and generally stable NSs) should be different as an additional condition $\d M_{\rm{NS}}/\d R=0$ is required for the former. 
In particular, only at the cores of TOV NSs the condition $\d M_{\rm{NS}}/\d\varepsilon_{\rm{c}}=0$ is required. 
Using the latter for TOV NSs via the mass $M_{\rm{NS}}$ scaling of Eq.\,(\ref{sca-M}), we obtained their central speed of sound squared (SSS) as
\begin{equation}\label{sc2-kk}
s_{\rm{c}}^2
\equiv\left.\frac{\d P}{\d \varepsilon}\right|_{\rm{center}}
=\x\left(1+\frac{1}{3}\frac{1+3\x^2+4\x}{1-3\x^2}\right)\left(1-\frac{3}{25}\x\right),
\end{equation}
see Appendix \ref{app-correction} for details of its derivation. With the stiffest possible EOS, $s_{\rm{c}}^2=1$ to remain casual. Considering the correction ``$-3\x/25$'', by setting $s_{\rm{c}}^2=1$ the $\x$ is now upper bounded by causality as $\x\lesssim0.381$ instead of $0.374$ previously obtained from the original mass scaling without its high-order correction ``$18\rm{X}/25$''\,\cite{CLZ23-a}.
In the following, we employ both the revised and original mass and compactness scalings, with and without the high-order correction terms, respectively, and then compare their outcomes.
For non-TOV NSs, the $1/3$ factor in Eq.\,(\ref{sc2-kk}) should be replaced by $(1+\Psi)/3$\,\cite{CLZ23-b} with $\Psi\equiv 2\d\ln M_{\rm{NS}}/\d\ln\varepsilon_{\rm{c}}>0$; therefore the $\x$ limited by $s_{\rm{c}}^2\leq1$ is smaller than 0.381. Since the TOV NSs are the most compact, the $\xi$ and correspondingly the $\x$ are the maximum.

\renewcommand{\arraystretch}{1.}
\renewcommand*\tablename{\small TAB.}
\begin{table}[h!]
\centerline{\begin{tabular}{c|c|c} 
  \hline\hline
&[$\overline{\Gamma}_{\rm{c}}$, $\overline{\Pi}_{\rm{c}}$]&[$\Gamma_{\rm{c}}$, $\Pi_{\rm{c}}$]\\\hline\hline
coefficient $A_{\xi}$&$1.56_{-0.016}^{+0.016}$&$2.31_{-0.03}^{+0.03}$\\\hline
coefficient $B_{\xi}$&$0.02_{-0.002}^{+0.002}$&$-0.032_{-0.003}^{+0.003}$\\\hline
coefficient $A_{\rm{M}}$&$865_{-8}^{+8}$&$1242_{-15}^{+15}$\\\hline
coefficient $B_{\rm{M}}$&$0.265_{-0.014}^{+0.014}$&$-0.08_{-0.02}^{+0.02}$\\\hline\hline
r-value for $\xi$-scaling&0.975&0.972\\\hline
r-value for $M_{\rm{NS}}$-scaling&0.978&0.965\\\hline
ste for $\xi$-scaling&0.02&0.03\\\hline
ste for $M_{\rm{NS}}$-scaling&$0.001M_{\odot}$&$0.002M_{\odot}$\\\hline\hline
$\xi_{\rm{GR}}$ (upper bound)&$0.276_{-0.003}^{+0.003}$&$0.264_{-0.005}^{+0.005}$\\\hline
$\Delta_{\rm{GR}}\leftrightarrow\x\lesssim3^{-1}-\Delta_{\rm{GR}}$&$-0.048$&$-0.041$\\\hline\hline
    \end{tabular}}
        \caption{Fitting properties of the NS mass and NS compactness scalings under two schemes. The coefficients $A_\xi$ and $B_\xi$ appear in $\xi=A_\xi\Pi_{\rm{c}}+B_\xi$ or $\xi=A_\xi\overline{\Pi}_{\rm{c}}+B_\xi$, while the coefficients $A_{\rm{M}}$ and $B_{\rm{M}}$ appear in $M_{\rm{NS}}/M_{\odot}=A_{\rm{M}}\Gamma_{\rm{c}}+B_{\rm{M}}$ or $M_{\rm{NS}}/M_{\odot}=A_{\rm{M}}\overline{\Gamma}_{\rm{c}}+B_{\rm{M}}$.
        }\label{tab_X1}        
\end{table}

The validity of the above scalings can be verified by comparing the results with those obtained by solving exactly the original TOV equations. As we shall show, using $10^5$ randomly generated NS meta-model EOSs satisfying all existing constraints from nuclear physics and astrophysics, the correction ``$18\x/25$'' is found to improve both the NS mass and compactness scalings for generally stable NSs along the mass-radius (M-R) curve. A NS EOS meta-model is a template for building models that can mimic most if not all  existing NS EOSs in the literature\,\cite{ZLX,ZLi-2,XL,ZLi-3}. Moreover, in order to verify the scalings in a broad sense, some of the EOS parameters are purposely set to be outside their currently known empirical ranges to generate some ``rare'' or ``unrealistic'' EOSs; see Appendix \ref{app-meta} for a detailed description on the meta-model EOS.

To explore the whole EOS parameter space and present our results clearly, we select randomly one point on each M-R curve from a given EOS within the mass range of $1.2M_{\odot}\lesssim M_{\rm{NS}}\lesssim2.2M_{\odot}$. The resulting scalings are shown in FIG.\,\ref{fig_Pic}. In each panel, only 500 representative samples are shown as scatters while $10^5$ EOSs are used in calculating the scaling coefficients and their error bands.
Here the panel (a) shows the compactness-$\overline{\Pi}_{\rm{c}}$ scaling while the panel (b) is the mass-$\overline{\Gamma}_{\rm{c}}$ scaling. 
The standard error (ste) and the coefficient of determination (the r-value) actually start converging quickly using about 300 samples.

We found that the ste for the compactness-$\overline{\Pi}_{\rm{c}}$ (mass-$\overline{\Gamma}_{\rm{c}}$) regression is about 0.02 (0.001$M_{\odot}$) and the r-values for the two regressions are about 0.975 and 0.978, respectively.
The high r-values via these meta-model EOSs (including the ``unrealistic'' ones) together with those from EOSs having multiple peaks or discontinuities in $s^2$ imply that the resulting scalings are really the intrinsic properties of the TOV equations rather than those of the input EOSs.
The regression and its 68\% confidence interval (CI) are shown by the light-blue (tomato) line and lavender (pink) band, respectively. Quantitatively, we have $\xi\approx A_\xi\overline{\Pi}_{\rm{c}}+B_\xi\approx1.56_{-0.016}^{+0.016}\overline{\Pi}_{\rm{c}}+0.02_{-0.002}^{+0.002}$ for the compactness-$\overline{\Pi}_{\rm{c}}$ scaling and $M_{\rm{NS}}/M_{\odot}\approx A_{\rm{M}}\overline{\Gamma}_{\rm{c}}+B_{\rm{M}}\approx865_{-8}^{+8}\overline{\Gamma}_{\rm{c}}+0.265_{-0.014}^{+0.014}$ for the mass-$\overline{\Gamma}_{\rm{c}}$ scaling, respectively.
If we adopt the original scalings $M_{\rm{NS}}\propto\Gamma_{\rm{c}}$ and $\xi\propto\Pi_{\rm{c}}$, the overall fitting results become less accurate, e.g., the r-values are 0.965 and 0.972, respectively. A detailed numerical comparison is provided in TAB.\,\ref{tab_X1}. For a graphical comparison, the original compactness and mass scalings without the correction $18\x/25$ are shown in the lower two panels ((a$'$) and (b$'$)) of FIG.\,\ref{fig_Pic}.
Compared to the original compactness scaling, it is seen that the revised one improves the linearity; while the revised mass scaling exhibits a smaller spread.
Moreover, since $\x$ is limited to $\x\lesssim0.381$ by causality realized in General Relativity (GR) with strong-field gravity, we have $\xi\lesssim0.276_{-0.003}^{+0.003}\equiv \xi_{\rm{GR}}$ based on the above compactness-$\overline{\Pi}_{\rm{c}}$ scaling. This result is consistent with the upper bound 0.33 for $\xi$ extracted in Ref.\,\cite{Ann22}. 

We also analyzed the radius-$\nu_{\rm{c}}$ correlation and obtained the scaling $R/\rm{km}\approx A_{\rm{R}}\nu_{\rm{c}}+B_{\rm{R}}\approx 572_{-25}^{+25}\nu_{\rm{c}}+4.22_{-0.35}^{+0.35}$ with a r-value about 0.712 and the $R$-ste of about 0.3\,km which is much smaller than that in current NS observations. As noticed earlier, dividing $R$ by $M_{\rm{NS}}$ in calculating the compactness $\xi$ largely diminishes the relatively large uncertainty in the radius scaling.

Shown also in panel (a) of FIG.\,\ref{fig_Pic} are the compactnesses for PSR J0740+6620 and PSR J0030+0451 via NICER's simultaneous mass-radius observation, namely $M_{\rm{NS}}/M_{\odot}\approx2.08_{-0.07}^{+0.07}$ and $R/\rm{km}\approx12.39_{-0.98}^{+1.30}$ (at 95\% CI) for the former\,\cite{Riley21} and $M_{\rm{NS}}/M_{\odot}\approx1.34_{-0.16}^{+0.15}$ and $R/\rm{km}\approx12.71_{-1.19}^{+1.14}$ (at 68\% CI) for the latter\,\cite{Riley19}, both indicated by the superscript ``a''.
Additionally, the compactness $\xi\approx0.183\sim0.259$ for the NS in GS 1826-24 directly from its surface gravitational red-shift measurement\,\cite{Zhou23} and the $\xi$ for a canonical NS with $R\approx12_{-1}^{+1}\,\rm{km}$\,\cite{Brandes23-a,Rich23} are also shown. For a given $\xi$, one can directly obtain the $\overline{\Pi}_{\rm{c}}$ from their scaling and the $\x$ via the function $\x(\overline{\Pi}_{\rm{c}})$ (orange dashed line) defined in Eq.\,(\ref{sca-xi}).

Similarly, the mass bands for PSR J0740+6620\,\cite{Fon21} and PSR J0030+0451\,\cite{Riley19} are shown in panel (b).
Given a mass $M_{\rm{NS}}$, the mass-$\overline{\Gamma}_{\rm{c}}$ scaling upper bounds the central energy density $\varepsilon_{\rm{c}}$ allowed since $\x\lesssim0.381$. For example, for a canonical NS, this upper bound is about $\y\equiv \varepsilon_{\rm{c}}/\varepsilon_0\lesssim13.7$ where $\varepsilon_0\approx150\,\rm{MeV}/\rm{fm}^3$ is the energy density of symmetric nuclear matter at its saturation density $\rho_0$. While for a massive NS of mass $M_{\rm{NS}}/M_{\odot}=2$ or $M_{\rm{NS}}/M_{\odot}=2.3$, we have $\y\lesssim5.7$ or $\y\lesssim4.2$, respectively.
We emphasize that this does not imply a canonical NS necessarily possesses a reduced central energy density of approximately 13.7; instead this value represents the upper limit allowed. By considering this counter-intuitive feature of NSs (see, e.g., Ref.\,\cite{SL24}) alongside the stability condition $\mathrm{d}M_{\rm NS}/\mathrm{d}\varepsilon_{\rm c} > 0$, one can gain physical insight into why a maximum mass for stable NSs must exist\,\cite{CL25}.

\section{Extracting NS Observational Constraints on Central Trace Anomalies Insensitive of EOS Models}\label{sec_ta}

The strong linear correlation between $\xi$ and $\overline{\Pi}_{\rm{c}}$ of Eq.\,(\ref{sca-xi}) enables us to read off the $\x$ straightforwardly from the $\xi$ obtained either using NS mass-radius observation or the red-shift measurement; and therefore the central trace anomaly $\Delta_{\rm{c}}=1/3-\x$.
Since the ratio $\phi=P/\varepsilon$ is an increasing function of $\varepsilon$ near NS centers, $\phi$ reaches its local maximum value $\x$ there, where the trace anomaly takes its local minimum $\Delta_{\rm{c}}$. This means $\phi\lesssim\x$ near the NS centers. We show it generally in our approach by anatomizing the structures of the TOV equations\,\cite{CL25}:
\begin{equation}
    \phi\approx\x-\frac{1}{6}\frac{1+\Psi}{4+\Psi}\left(1+\frac{7+\Psi}{4+\Psi}\cdot 4\x\right)\widehat{r}^2\to\x-\frac{1+7\x}{24}\widehat{r}^2,
\end{equation}
where $\widehat{r}$ is the dimensionless radial distance from the center; the second relation follows for TOV NSs ($\Psi=0$).
A lower bound for trace anomaly $\Delta\gtrsim \Delta_{\rm{GR}}\approx-0.048$ from the GR limit $\x\lesssim0.381$ is equivalent to $\xi\lesssim0.276_{-0.003}^{+0.003}$ for the compactness discussed earlier.

We notice that in a recent study\,\cite{Ecker22-a}, the minimum of $\Delta$ is found to be slightly away from the NS center when incorporating constraints on $\Delta$ at extremely high densities from pQCD theories. On the other hand, the state-of-the-art calculation indicates that $\Delta$ likely decreases all the way to $M_{\rm{TOV}}$, as shown by the lower-right panel of figure 2 of Ref.\,\cite{Ann23}.
In fact, the position of a local minimum of $\Delta$ (if it exists) being smaller than $\varepsilon_{\rm{c}}$ has a strong implication on the existence of a possible peak in the density profile of $s^2$ within NS densities. However, the inverse is not generally true: If there is a peak in $s^2$ within NS densities, the $\Delta$ within NS densities may or may not develop a local minimum, we prove this statement in the last paragraph of Appendix \ref{app-s2}.

\begin{figure*}
\centering
\includegraphics[width=7.2cm]{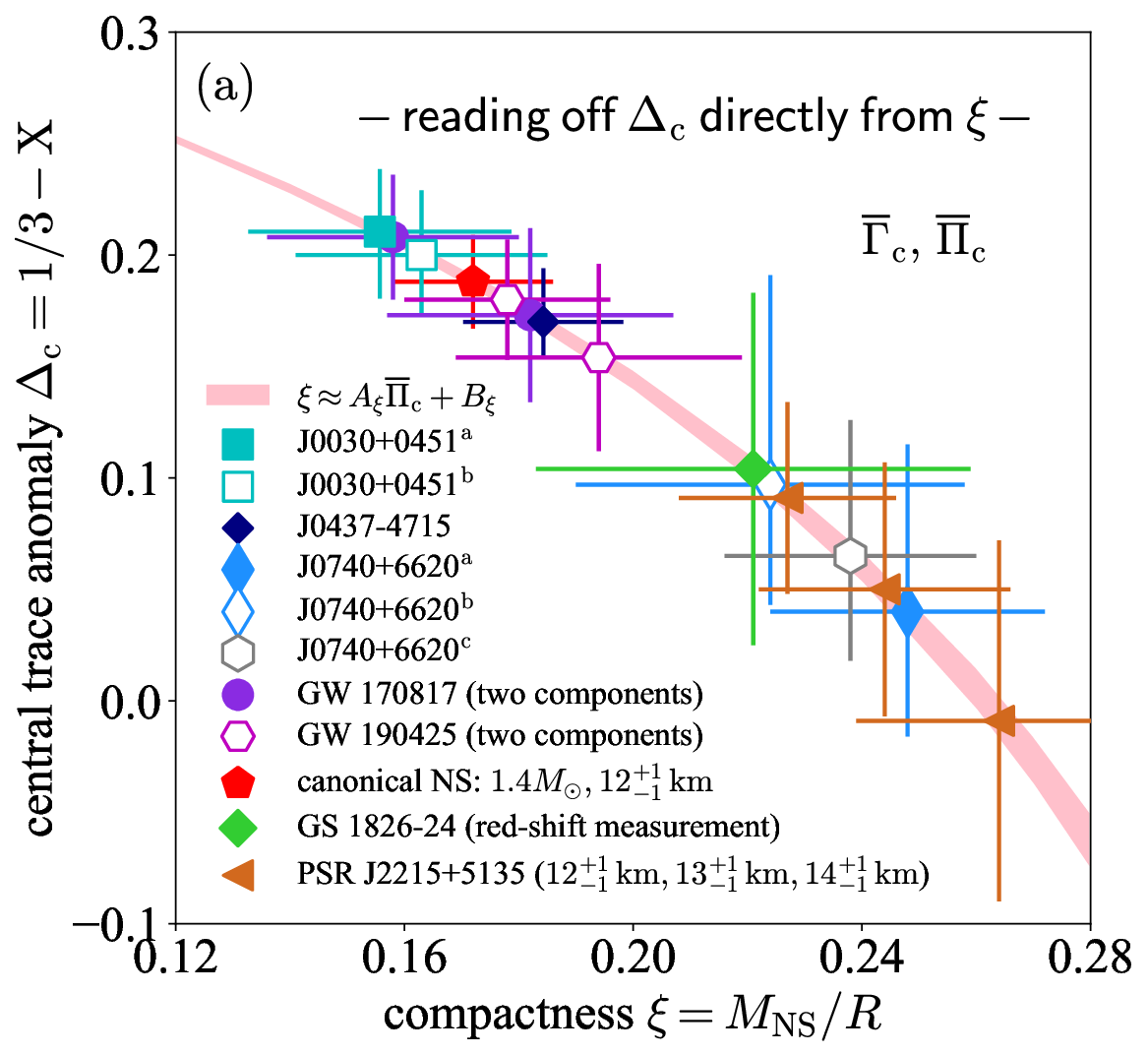}\qquad
\includegraphics[width=7.cm]{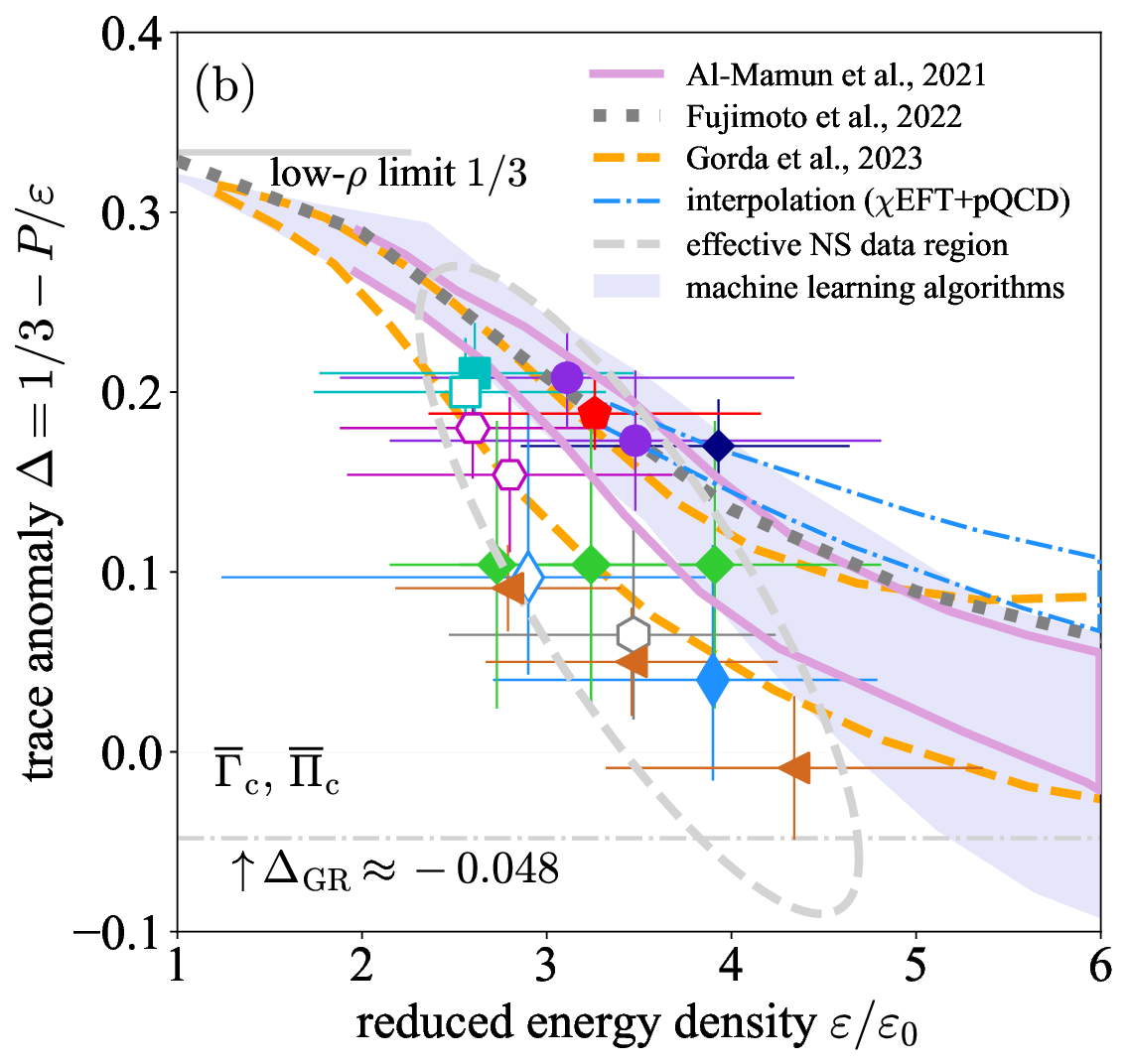}\\[0.25cm]
\includegraphics[width=7.2cm]{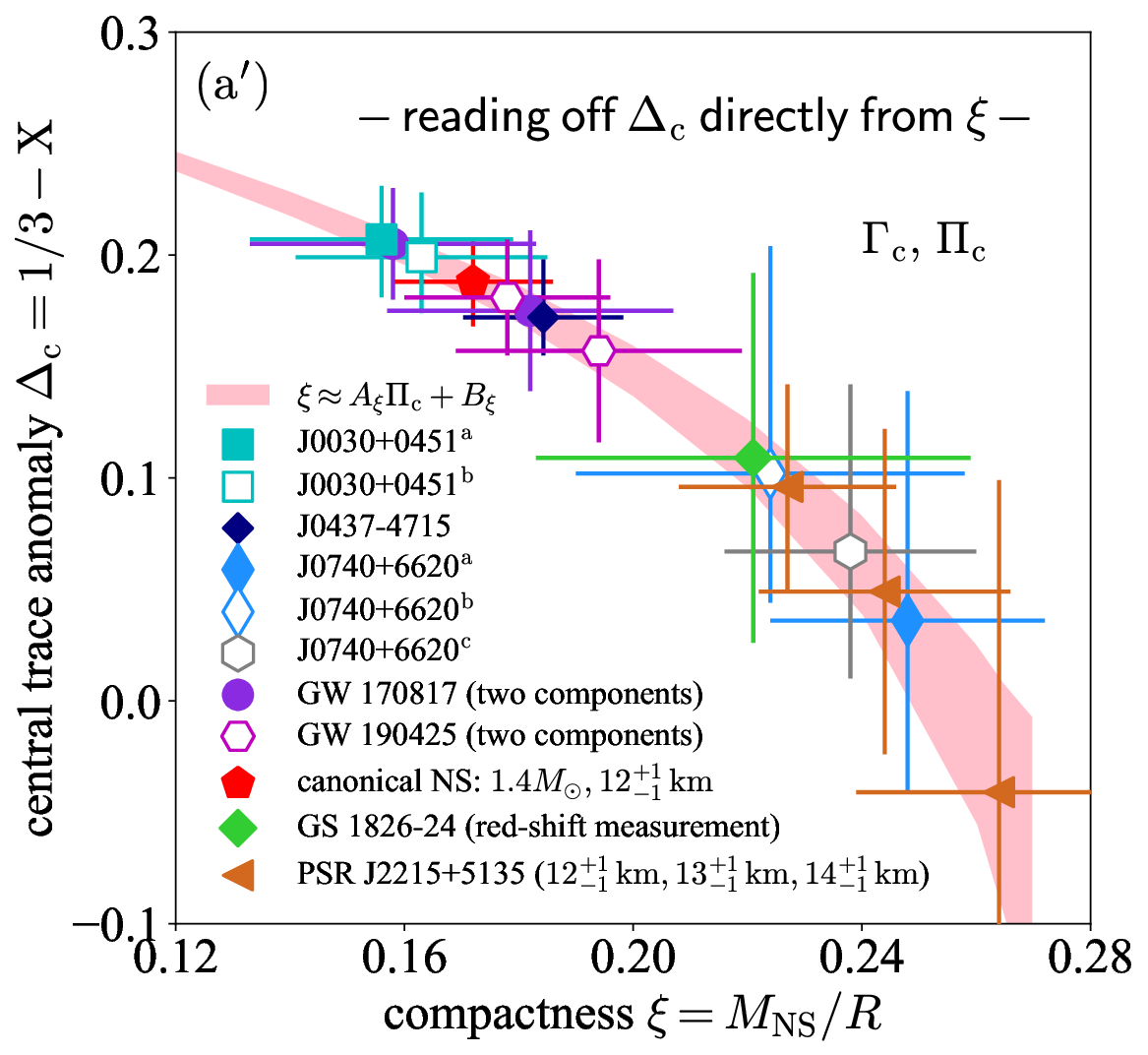}\qquad
\includegraphics[width=7.cm]{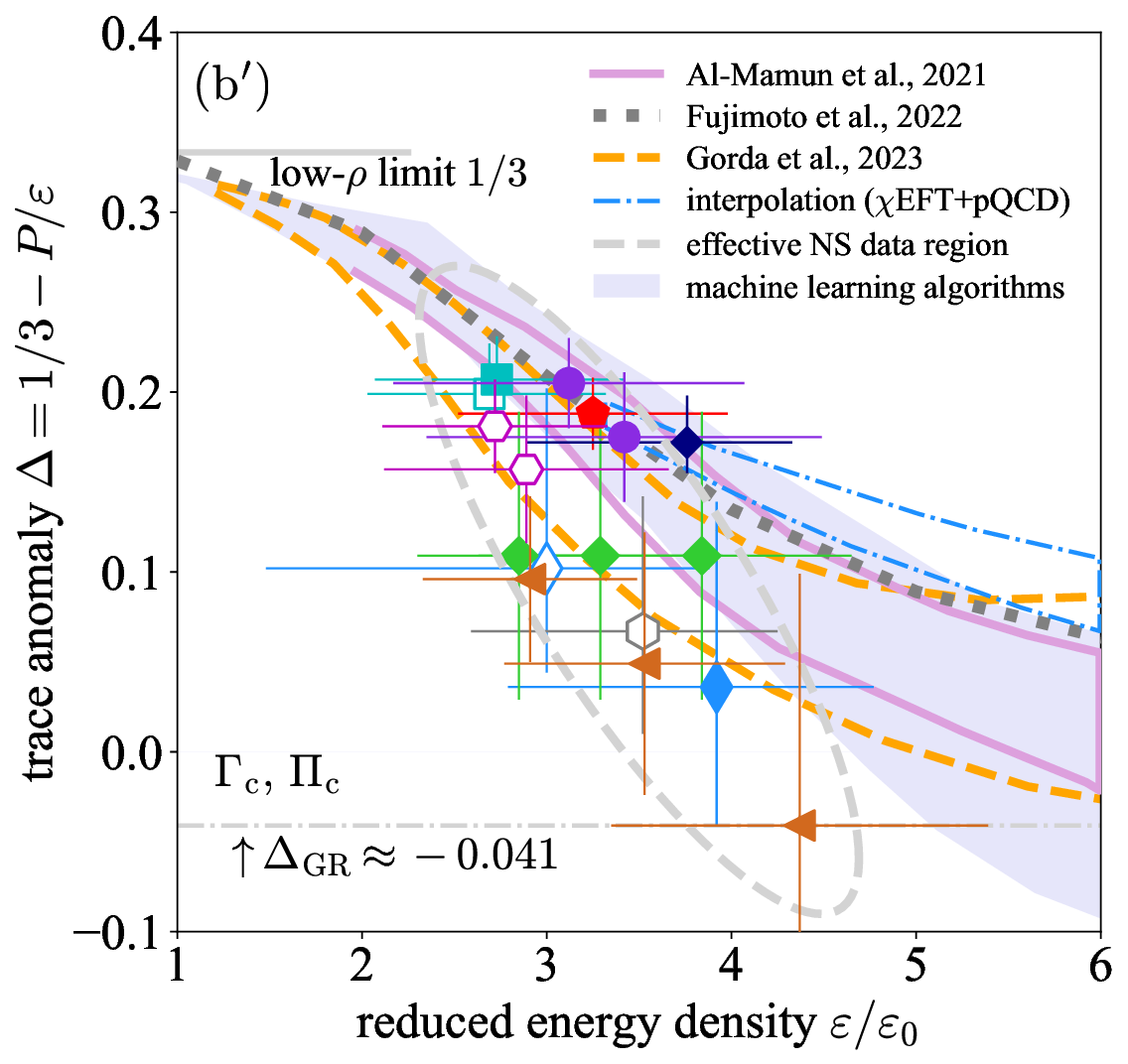}
\caption{(Color Online).  Upper left panel: Central trace anomaly $\Delta_{\rm{c}}$ as a function of compactness $\xi$ from inverting the compactness scaling (pink band) in comparison with observational data indicated. Upper right panel: Energy density dependence of the trace anomaly where the trace anomalies from a few empirical NS EOSs via different input data/inference algorithms are also given. The GR bound $\Delta\gtrsim \Delta_{\rm{GR}}\approx-0.048$ is plotted by the grey dash-dotted line, see  text for more details.
The errors are at $1\sigma$ level.
Lower two panels: The same as the upper two panels but using the original compactness and mass scalings.
}\label{fig_Delta}
\end{figure*}

\renewcommand{\arraystretch}{1.}
\begin{table}[b!]
\centerline{\begin{tabular}{c|c|c||c|c} 
  \hline\hline
NS&$\Delta_{\rm{c}}$ [$\overline{\Gamma}_{\rm{c}}$, $\overline{\Pi}_{\rm{c}}$]&$\y$ [$\overline{\Gamma}_{\rm{c}}$, $\overline{\Pi}_{\rm{c}}$]&$\Delta_{\rm{c}}$ [${\Gamma}_{\rm{c}}$, ${\Pi}_{\rm{c}}$]&$\y$ [${\Gamma}_{\rm{c}}$, ${\Pi}_{\rm{c}}$]\\\hline\hline
0030+0451$^{\rm{a}}$&$0.211_{-0.030}^{+0.028}$&$2.61_{-0.84}^{+0.86}$&$0.207_{-0.026}^{+0.024}$&$2.73_{-0.66}^{+0.69}$\\\hline
0030+0451$^{\rm{b}}$&$0.200_{-0.027}^{+0.029}$&$2.56_{-0.82}^{+0.76}$&$0.199_{-0.025}^{+0.029}$&$2.69_{-0.66}^{+0.63}$\\\hline
0437-4715&$0.170_{-0.016}^{+0.024}$&$3.93_{-1.07}^{+0.71}$&$0.172_{-0.017}^{+0.026}$&$3.76_{-0.87}^{+0.57}$\\\hline
0740+6620$^{\rm{a}}$&$0.040_{-0.057}^{+0.075}$&$3.90_{-1.19}^{+0.89}$&$0.036_{-0.077}^{+0.103}$&$3.92_{-1.13}^{+0.85}$\\\hline
0740+6620$^{\rm{b}}$&$0.097_{-0.054}^{+0.094}$&$2.90_{-1.66}^{+0.96}$&$0.102_{-0.058}^{+0.102}$&$3.00_{-1.52}^{+0.88}$\\\hline
0740+6620$^{\rm{c}}$&$0.065_{-0.047}^{+0.061}$&$3.47_{-1.00}^{+0.77}$&$0.067_{-0.057}^{+0.075}$&$3.52_{-0.93}^{+0.73}$\\\hline
GW 170817-1&$0.173_{-0.039}^{+0.039}$&$3.48_{-1.33}^{+1.33}$&$0.175_{-0.036}^{+0.036}$&$3.42_{-1.07}^{+1.07}$\\\hline
GW 170817-2&$0.208_{-0.028}^{+0.028}$&$3.11_{-1.23}^{+1.23}$&$0.205_{-0.025}^{+0.025}$&$3.12_{-0.95}^{+0.95}$\\\hline
GW 190425-1&$0.154_{-0.041}^{+0.041}$&$2.80_{-0.88}^{+0.88}$&$0.157_{-0.040}^{+0.040}$&$2.89_{-0.77}^{+0.77}$\\\hline
GW 190425-2&$0.180_{-0.027}^{+0.027}$&$2.60_{-0.72}^{+0.72}$&$0.181_{-0.025}^{+0.025}$&$2.72_{-0.61}^{+0.61}$\\\hline
canonical NS&$0.188_{-0.021}^{+0.021}$&$3.26_{-0.90}^{+0.90}$&$0.188_{-0.020}^{+0.020}$&$3.25_{-0.73}^{+0.73}$\\\hline
GS 1826-24$^{\rm{a}}$&$0.104_{-0.079}^{+0.079}$&$3.91_{-0.90}^{+0.90}$&$0.109_{-0.082}^{+0.082}$&$3.84_{-0.81}^{+0.81}$\\\hline
GS 1826-24$^{\rm{b}}$&$0.104_{-0.079}^{+0.079}$&$3.24_{-0.71}^{+0.71}$&$0.109_{-0.082}^{+0.082}$&$3.29_{-0.66}^{+0.66}$\\\hline
GS 1826-24$^{\rm{c}}$&$0.104_{-0.079}^{+0.079}$&$2.73_{-0.58}^{+0.58}$&$0.109_{-0.082}^{+0.082}$&$2.85_{-0.55}^{+0.55}$\\\hline
2215+5135$^{\rm{a}}$&$-0.009_{-0.081}^{+0.081}$&$4.34_{-1.02}^{+1.02}$&$-0.041_{-0.140}^{+0.140}$&$4.37_{-1.02}^{+1.02}$\\\hline
2215+5135$^{\rm{b}}$&$0.050_{-0.057}^{+0.057}$&$3.46_{-0.79}^{+0.79}$&$0.049_{-0.073}^{+0.073}$&$3.53_{-0.76}^{+0.76}$\\\hline
2215+5135$^{\rm{c}}$&$0.091_{-0.043}^{+0.043}$&$2.79_{-0.61}^{+0.61}$&$0.096_{-0.046}^{+0.046}$&$2.91_{-0.58}^{+0.58}$\\\hline\hline
    \end{tabular}}
        \caption{Central dimensionless trace anomaly $\Delta_{\rm{c}}$ and the central energy density $\y=\varepsilon_{\rm{c}}/\varepsilon_0$ (reduced by $\varepsilon_0\approx150\,\rm{MeV}/\rm{fm}^3$) of the 17 NS instances under two scaling schemes; here ``$0030+0451^{\rm{a,b}}$'' uses radius from Refs.\,\cite{Riley19,Miller19}, ``$0740+6620^{\rm{a,b,c}}$'' uses radius from Refs.\,\cite{Riley21,Miller21,Salmi22}, the radii for GS 1826-24$^{\rm{a,b,c}}$ and PSR J2215+5135$^{\rm{a,b,c}}$ are assumed to be $12_{-1}^{+1}\,\rm{km}$, $13_{-1}^{+1}\,\rm{km}$ and $14_{-1}^{+1}\,\rm{km}$, respectively. See text for details.
        }\label{tab_X2}        
\end{table}

In panel (a) of FIG.\,\ref{fig_Delta}, we show the $\xi$-dependence of $\Delta_{\rm{c}}$ (with errorbars) by inverting $\xi\approx A_\xi\overline{\Pi}_{\rm{c}}(\x)+B_\xi$ (pink band).
Here 15 NS instances are shown,  these include two alternative inferences of the radius using somewhat different approaches for PSR J0030+0451\,\cite{Riley19,Miller19} by superscript ``a,b'' and three for PSR J0740+6620\,\cite{Riley21,Miller21,Salmi22} by ``a,b,c''; the
PSR J0437-4715 with its mass and radius about $1.418_{-0.037}^{+0.037}M_{\odot}$ and $11.36_{-0.63}^{+0.95}\,\rm{km}$\,\cite{Choud24} (see also Ref.\,\cite{Reardon24}); 
the NS in GS 1826-24\,\cite{Zhou23},  a canonical NS with radius $R\approx12_{-1}^{+1}\,\rm{km}$\,\cite{Brandes23-a}; and three central trace anomalies for PSR J2215+5135\,\cite{Sul24}. 
Though there is currently no observational constraint on the radius of PSR J2215+5135, we can use the $\xi_{\rm{GR}}$ or $\Delta_{\rm{GR}}$ to limit $R\gtrsim11.5\,\rm{km}$. Here three typical radii, namely $12_{-1}^{+1}$\,km, $13_{-1}^{+1}$\,km and $14_{-1}^{+1}$\,km are adopted for an illustration. It is known that the two NSs involved in GW 170817\,\cite{Abbott2018} have $R_1\approx R_2\approx11.9_{-1.4}^{+1.4}\,\rm{km}$ as well as $1.36\lesssim M_{\rm{NS1}}/M_{\odot}\lesssim1.58$ and $1.18\lesssim M_{\rm{NS2}}/M_{\odot}\lesssim1.36$, respectively;
while GW 190425\,\cite{Abbott2020-a} has $12.0\,\rm{km}\lesssim R_1\lesssim14.6\,\rm{km}$ with $1.6\lesssim M_{\rm{NS1}}/M_{\odot}\lesssim1.9$ and $12.2\,\rm{km}\lesssim R_2\lesssim14.4\,\rm{km}$ with $1.5\lesssim M_{\rm{NS2}}/M_{\odot}\lesssim1.7$, respectively, using the low-spin prior\,\cite{Abbott2021}.
The error bar of $\Delta_{\rm{c}}$ (at $1\sigma$ level) is mainly due to the uncertainty of $\xi$ itself as the correlation $\xi$-$\overline{\Pi}_{\rm{c}}$ is strong and model independent.
For instance,  the error bar of $\Delta_{\rm{c}}$ for a canonical NS with $R\approx12_{-1}^{+1}\,\rm{km}$ is apparently smaller than that for the two NSs in GW 170817\,\cite{Abbott2018} as they have larger mass uncertainties although they share similar radii (compare the red solid pentagon and dark-violet solid circles).

Besides the $\xi$-$\overline{\Pi}_{\rm{c}}$ scaling, the mass-$\overline{\Gamma}_{\rm{c}}$ scaling of Eq.\,(\ref{sca-M}) further gives individually the values of $P_{\rm{c}}$ and $\varepsilon_{\rm{c}}$ if both the $M_{\rm{NS}}$ and $R$ (or one of them together with the compactness) are observationally known. 
In order to obtain $\x$ and $\varepsilon_{\rm{c}}$ for the NS in GS 1826-24 (only its compactness is known), we adopt three typical radii ($12_{-1}^{+1}$\,km, $13_{-1}^{+1}$\,km and $14_{-1}^{+1}$\,km), the same as that for PSR J2215+5135.
The numerical values of $\Delta_{\rm{c}}$ and $\y\equiv\varepsilon_{\rm{c}}/\varepsilon_0$ for the 17 NS instances (including the previous 15 NSs shown in panel (a) of FIG.\,\ref{fig_Delta} and the extra two for the NS in GS 1826-24) are given in TAB.\,\ref{tab_X2}. They are also displayed in panel (b) of FIG.\,\ref{fig_Delta} and enclosed by the dashed grey ellipse (the effective region of NS data). For comparisons, also shown are the trace anomalies obtained/constrained by a few contemporary state-of-the-art NS EOS modelings using different input data and/or inference algorithms. These include the NS EOS inference\,\cite{Gorda23} incorporating the pQCD impact (dashed orange band), a Bayesian inference of NS EOS\,\cite{Mam2021} combining the electromagnetic and gravitational-wave signals (plum solid band), the interpolation\,\cite{Fuji22} between low-density chiral effective field theories\,\cite{Ess21} (CEFT) and high-density pQCD constraints\,\cite{Bjorken83,Kurkela10,Gorda21PRL,Gorda23PRL} (dash-dotted light-blue band),  a minimal parametrization\,\cite{Fuji22} of $\Delta$ versus $\varepsilon/\varepsilon_0$ (grey dotted line) accounting for NS data and the NS EOS\,\cite{Fuji22} inferred via machine learning algorithms (lavender band).

\begin{figure}[h!]
\centering
\includegraphics[width=8.5cm]{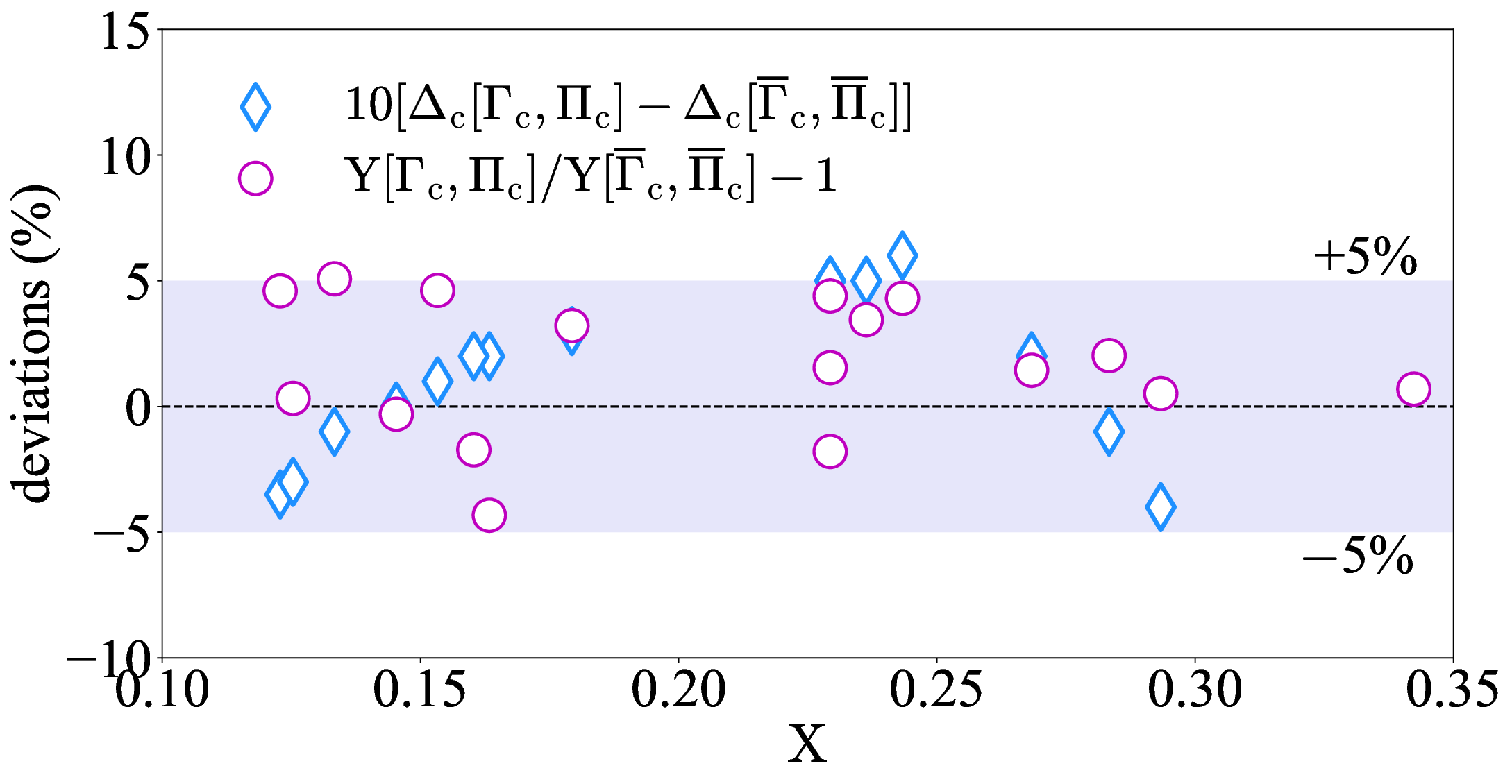}
\caption{(Color Online). Deviations of the trace anomaly and the reduced central energy density in the two scaling schemes.
}\label{fig_deltaY_on_X}
\end{figure}

Similarly, the extracted values of $\Delta_{\rm{c}}$ and $\y$ using the original compactness and mass scalings for these 17 NS instances are shown in the lower two panels ((a$'$) and (b$'$)) of FIG.\,\ref{fig_Delta} and in TAB.\,\ref{tab_X2}. We observe that the $\y$ extracted from both scaling schemes exhibit similar uncertainties,  e.g., both the $\Delta_{\rm{c}}$ and $\y$ for a canonical NS with $R\approx12_{-1}^{+1}\,\rm{km}$ are very close to each other; while the $\Delta_{\rm{c}}$ obtained using the revised scalings including the correction $18\x/25$ has smaller uncertainty for massive NSs.
Moreover, the overall spreading of the compactness-trace anomaly correlation in the revised scaling scheme is much smaller, by comparing panels (a) and (a$'$) of FIG.\,\ref{fig_Delta}.
In order to visualize the similarity between results in the two scaling schemes, we show in FIG.\,\ref{fig_deltaY_on_X} the deviation for $\Delta_{\rm{c}}$ using $10[\Delta_{\rm{c}}[\Gamma_{\rm{c}},\Pi_{\rm{c}}]-\Delta_{\rm{c}}[\overline{\Gamma}_{\rm{c}},\overline{\Pi}_{\rm{c}}]]$ as a metric
since the $\Delta_{\rm{c}}$ may be close to zero and that for $\y$ via $
\y[\Gamma_{\rm{c}},\Pi_{\rm{c}}]/\y[\overline{\Gamma}_{\rm{c}},\overline{\Pi}_{\rm{c}}]-1$.
It is seen that the magnitude of the absolute difference in $\Delta_{\rm{c}}$ from the two scaling schemes is generally smaller than 0.006 (light-blue diamonds); and the relative difference of $\y$ is smaller than about 5\% (magenta circles).
These results show that although the two scaling schemes perform slightly differently in the test using the $10^5$ NS meta-model EOS samples, the extracted $\Delta_{\rm{c}}$ and $\y$ from the NS observational data are rather robust.

Our results of $\y\equiv\varepsilon_{\rm{c}}/\varepsilon_0$ and $\Delta_{\rm{c}}=1/3-\x$ for the 17 NS instances put stringent constraints on the theoretical NS EOSs. 
As shown in panel (b) of FIG.\,\ref{fig_Delta}, apparently the $\Delta$'s and especially their energy dependence (which determines the speed of sound as we shall discuss next)  from some NS EOS modelings have sizable tensions with the limits set by the observational data based on our scaling analyses, especially for massive NSs. It implies that some ingredients in modeling NS EOS may need to be revised. For example, the NS EOS model incorporating pQCD effect\,\cite{Gorda23} shown by the dashed orange band can well explain the $(\y,\Delta_{\rm{c}})$'s of PSR J0030+0451, GW 190425 and GS 1826-24 (with a radius $R\approx12\,\rm{km}$ or 13\,km,  the rightest two green diamonds). However, it could hardly account for the central values of the results for PSR J0740+6620 for all three radii\,\cite{Riley21,Miller21,Salmi22} and the two NSs in GW 170817\,\cite{Abbott2018} as well as those for PSR J2215+5135\,\cite{Sul24} (for all three $R$'s). Similarly, the NS EOS modeling with both the electromagnetic and gravitational-wave signals included\,\cite{Mam2021} (plum solid band) can effectively explain the data of GW 170817 and the canonical NS. But it has certain tensions with our results based on observations for PSR J0030+451 and PSR J0740+6620, GW 190425 and the redback spider pulsar PSR J2215+5135. Interestingly, although GW 190425 executes weaker limits on NS radii\,\cite{Abbott2020-a},  it effectively puts useful constraints on the $\Delta$. 
Moreover, the PSR J0437-4715\,\cite{Choud24,Reardon24} with a relative small radius (navy solid diamond) generates some strong challenges on the modeling of $\Delta$.
On the other hand, the interpolation\,\cite{Fuji22} between low-density CEFT\,\cite{Ess21} and high-density pQCD\,\cite{Bjorken83,Kurkela10,Gorda21PRL,Gorda23PRL} (dash-dotted light-blue band) predicts a quite large $\Delta$ compared with what we extracted from PSR J0740+6620, GS 1826-24 and PSR J2215+5135 observations.

\begin{figure*}
\centering
\includegraphics[width=7.cm]{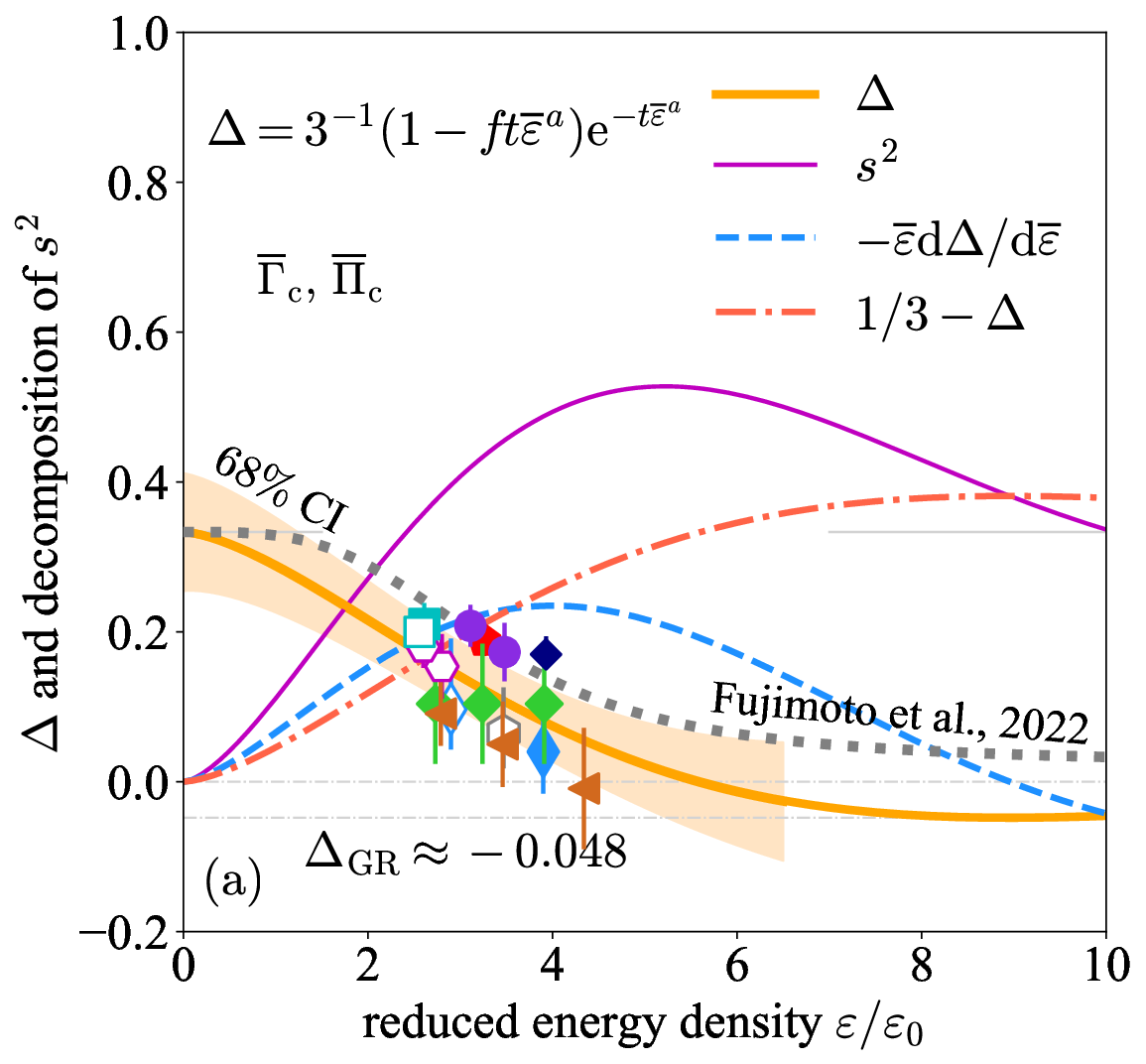}\qquad
\includegraphics[width=7.cm]{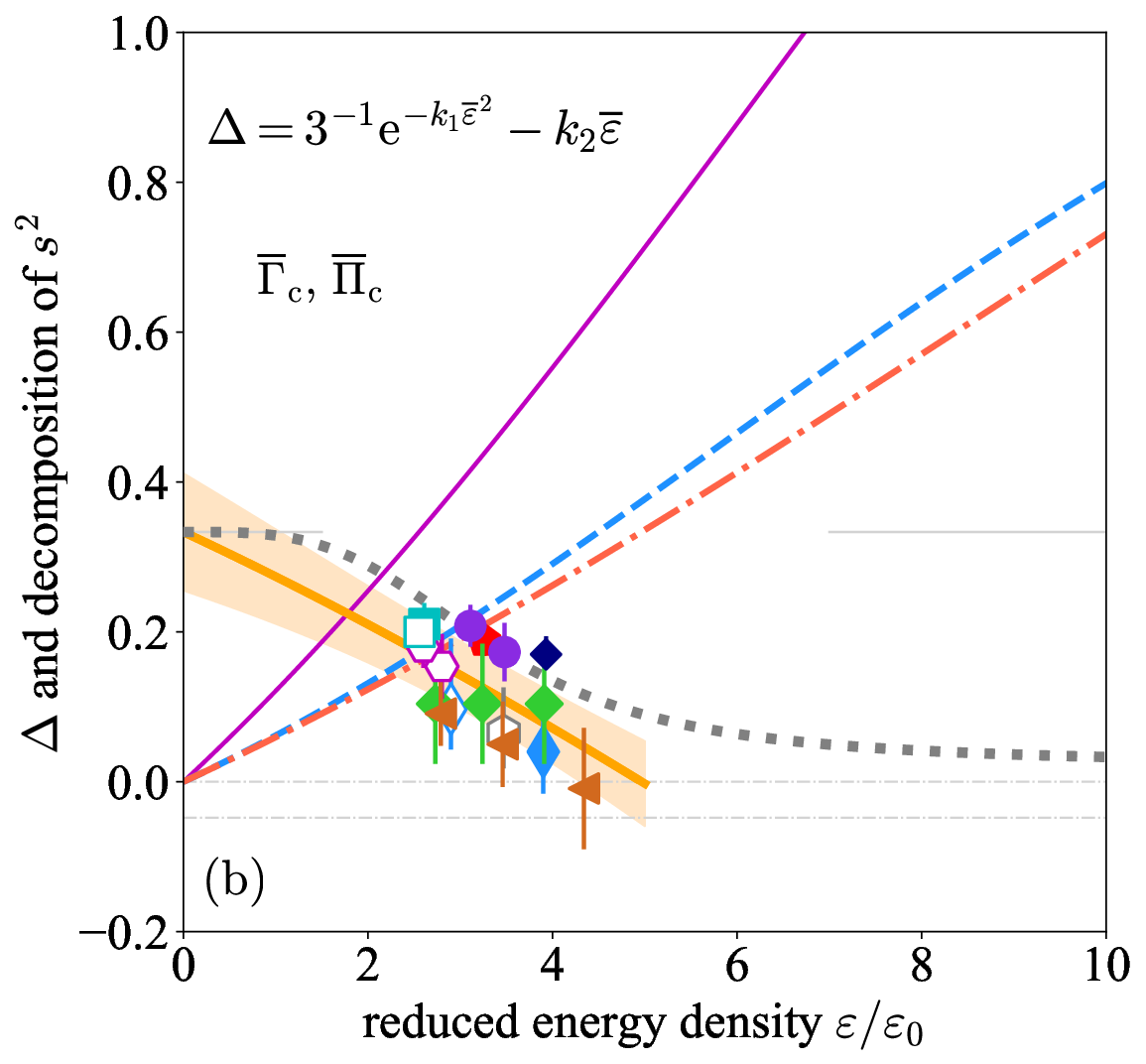}\\[0.25cm]
\includegraphics[width=7.cm]{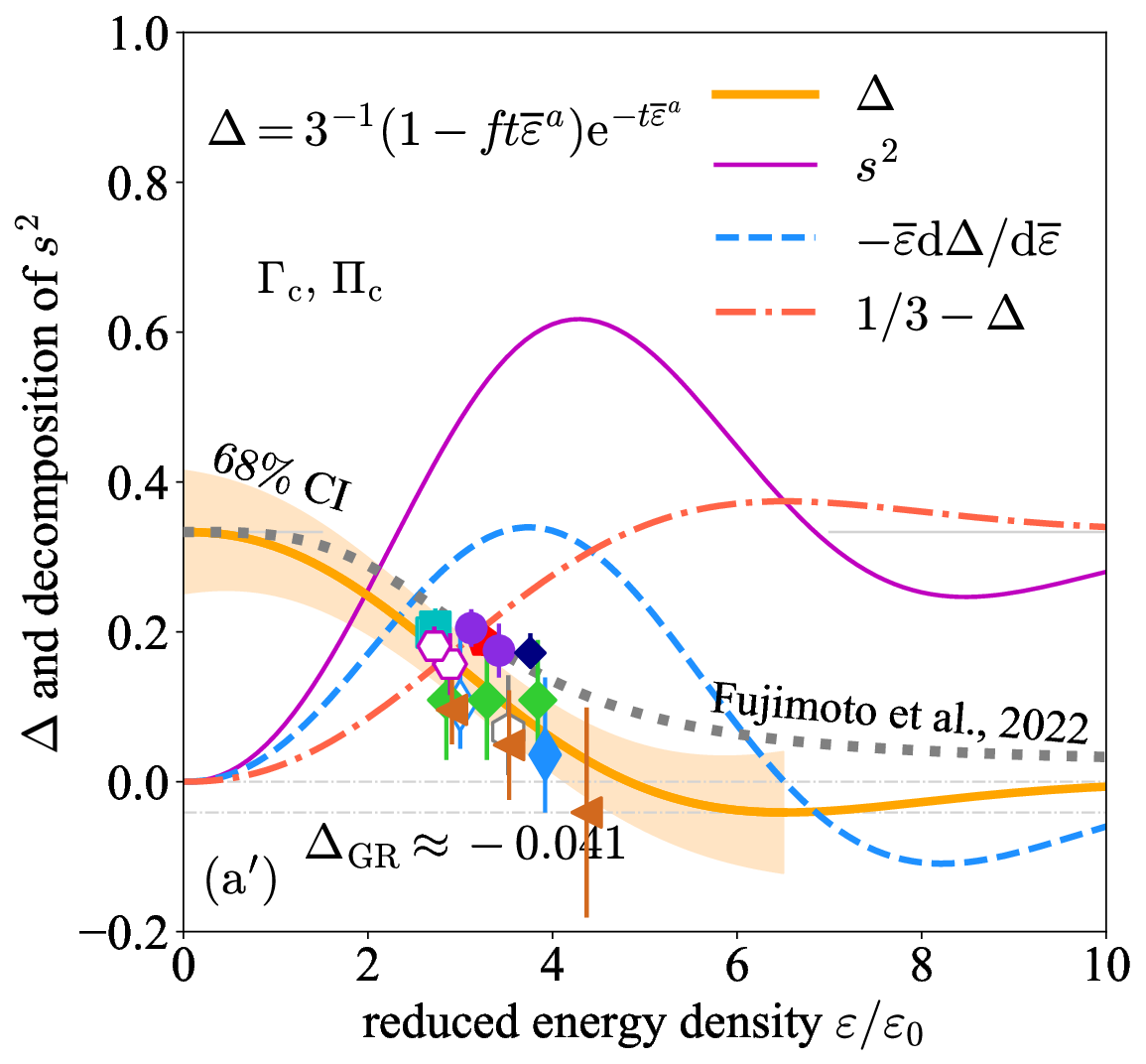}\qquad
\includegraphics[width=7.cm]{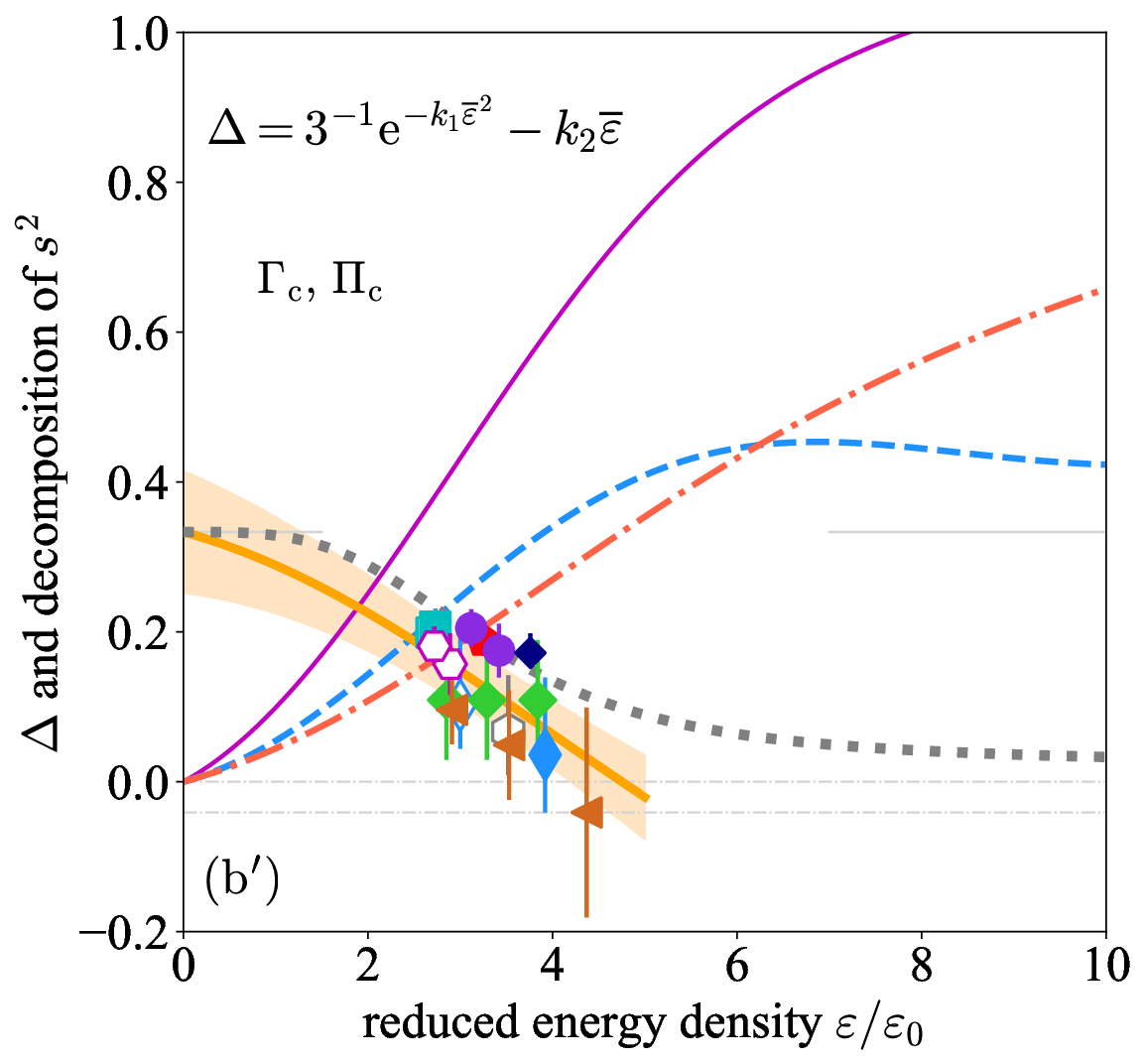}
\caption{(Color Online).  Upper two panels: The trace anomaly $\Delta$, $s^2$ and its decomposition via Eq.\,(\ref{ss_decom}) using two effective parametrizations for $\Delta$, in the revised scaling scheme.
The 17 NS instances of panel (b) of FIG.\,\ref{fig_Delta} are plotted with the tan band for its 68\% CI regression in each panel. The grey dotted line represents the minimal parametrization of Ref.\,\cite{Fuji22}.
Lower two panels: The same as the upper two panels but using the original compactness and mass scalings.
}\label{fig_s2ab}
\end{figure*}

\section{Do Existing NS Data Require Inevitably a Peaked Density Profile for the Speed of Sound?}\label{sec_s2}
The answer to this question is strongly dependent on the energy dependence of the trace anomaly underlying the available NS observational data. Our results above from analyzing NS compactness and mass scalings can help address this issue. This is because the trace anomaly $\Delta$ and its derivative with respect to energy density are crucial for understanding the $s^2$ profile in NSs\,\cite{Fuji22},
\begin{equation}\label{ss_decom}
s^2\equiv{\d P}/{\d\varepsilon}=-\overline{\varepsilon}{\d\Delta}/{\d\overline{\varepsilon}}+{3}^{-1}-\Delta,
\end{equation}
where $\overline{\varepsilon}\equiv\varepsilon/\varepsilon_0$. The first term is obviously the derivative while the remainings together represent the non-derivative part of $s^2$.
In the following, two effective parametrizations for $\Delta$ are adopted for illustration:
\begin{align}
\mbox{p1}:\Delta\approx&3^{-1}\left(1-ft\overline{\varepsilon}^{a}\right)\exp\left(-t\overline{\varepsilon}^a\right);\label{pp1}\\
\mbox{p2}:\Delta\approx&3^{-1}\exp\left({-k_1\overline{\varepsilon}^{2}}\right)-k_2\overline{\varepsilon}.\label{pp2}
\end{align}

Panel (a) of FIG.\,\ref{fig_s2ab} shows the $\Delta$ of p1 as a function of 
$\overline{\varepsilon}$ (orange curve) to averagely account for the 17 NS instances in the revised scaling scheme, valid for $2.5\lesssim\overline{\varepsilon}\lesssim4.5$ (data available). Here $a\approx 1.51$ and $t\approx 0.07$ are two parameters, and $f\equiv \rm{L}_W^{-1}(-3^{-1}/\rm{e}\Delta_{\rm{GR}})\approx1.0318$ with $\Delta_{\rm{GR}}\approx-0.048$ and $\rm{L}_W(x)$ being the Lambert-$W$ function defined as the solution of the equation $w\exp(w)=x$.
This parametrization has the properties: (i) $\Delta\to0$ for $\overline{\varepsilon}\to\infty$\,\cite{Bjorken83,Kurkela10,Gorda21PRL,Gorda23PRL}, and (ii) $\Delta\to1/3$ as $\overline{\varepsilon}\to0$; it also generates a minimum $\Delta_{\min}=\Delta_{\rm{GR}}$ at $\overline{\varepsilon}_{\rm{GR}}=[(f+1)/ft]^{1/a}\approx9.0$ by construction (through the factor $f$).
With the parametrization p1, there would be a peak generated in $s^2$.
Numerically, we find $\Delta$ in panel (a) of FIG.\,\ref{fig_s2ab} drops quickly at roughly about $\overline{\varepsilon}\approx3.5$ from being $\gtrsim0.1$ to $\lesssim0.1$.
This feature actually strongly connects with the possible peaked behavior of the SSS\,\cite{Fuji22}.
The expected peak emerges at $\overline{\varepsilon}_{\rm{pk}}\approx5.2$ or $\varepsilon_{\rm{pk}}\approx780\,\rm{MeV}/\rm{fm}^3$ with $s^2_{\rm{pk}}\approx0.53$; this is because $\Delta\to \Delta_{\rm{GR}}$ at $\overline{\varepsilon}_{\rm{GR}}\approx9.0$ and thus a peak emerges (at $\overline{\varepsilon}_{\rm{deriv,pk}}\approx4.1$) in the derivative term ``$-\overline{\varepsilon}\d\Delta/\d\overline{\varepsilon}$''\,\cite{Fuji22}.
So the peak position in $s^2$ is larger than that in $s_{\rm{deriv}}^2$.
Notice also that $\overline{\varepsilon}_{\rm{pk}}\approx5.2$ may be somewhat larger than the central energy density realized in NSs listed in TAB.\,\ref{tab_X2}.
More interestingly, the GR bound $\Delta_{\min}=\Delta_{\rm{GR}}<0$ induces a valley in $s^2$ at $\overline{\varepsilon}_{\rm{vl}}\approx13.7$ with $s^2_{\rm{vl}}\approx0.27$, since $\Delta\to0>\Delta_{\rm{GR}}$ as $\varepsilon/\varepsilon_0\to\infty$ due to a valley appearing at about $\overline{\varepsilon}_{\rm{deriv,vl}}\approx13.2$ in the derivative part. However, these densities may safely exceed the one allowed in realistic NSs.
In this sense, the pQCD limit for $\Delta$ is the origin for the emergence of a valley in $s^2$ instead of a peak\,\cite{Fuji22}.
It also shows that the valley position in $s^2$ is slightly larger than that in $s^2_{\rm{deriv}}$.
Similarly,  if we adopt the scaling scheme based on $\Gamma_{\rm{c}}$ and $\Pi_{\rm{c}}$, then $t\approx0.03$ and $a\approx2.24$ would be obtained; the corresponding properties of the peak and valley of the SSS are summarized in TAB.\,\ref{tab_X3}.
The SSS together with its decomposition terms in the original scaling scheme using parametrization p1 are shown in the lower two panels of FIG.\,\ref{fig_s2ab}.

\renewcommand{\arraystretch}{1.}
\begin{table}[tbh]
\centerline{\begin{tabular}{c|c|c} 
  \hline\hline
&[$\overline{\Gamma}_{\rm{c}}$, $\overline{\Pi}_{\rm{c}}$]&[$\Gamma_{\rm{c}}$,  $\Pi_{\rm{c}}$]\\\hline\hline
$a$&1.51&2.24\\\hline
$t$&0.07&0.03\\\hline\hline
upper bound on $\x$&0.381&0.374\\\hline
$\Delta_{\rm{GR}}$&$-0.048$&$-0.041$\\\hline
$f=\rm{L}_W^{-1}(-3^{-1}/\rm{e}\Delta_{\rm{GR}})$&1.0318&0.9539\\\hline
$\overline{\varepsilon}_{\rm{GR}}=[(f+1)/ft]^{1/a}$&9.0&6.5\\\hline\hline
$\overline{\varepsilon}_{\rm{pk}}$&5.2&4.3\\\hline
$s_{\rm{pk}}^2$&0.53&0.62\\\hline
$\overline{\varepsilon}_{\rm{vl}}$&13.7&8.4\\\hline
$s_{\rm{vl}}^2$&0.27&0.25\\\hline\hline
$\overline{\varepsilon}_{\rm{deriv,pk}}$&4.1&3.7\\\hline
$s_{\rm{deriv,pk}}^2$&0.23&0.34\\\hline
$\overline{\varepsilon}_{\rm{deriv,vl}}$&13.2&8.2\\\hline
$s_{\rm{deriv,vl}}^2$&$-0.07$&$-0.11$\\\hline\hline
    \end{tabular}}
        \caption{Fitting properties of the parametrization p1 of (\ref{pp1}) under two scaling schemes.
        }\label{tab_X3}        
\end{table}

On the other hand,  using the parametrization p2 which considers only the constraint $\lim_{\overline{\varepsilon}\to0}\Delta=3^{-1}$ (without its large $\overline{\varepsilon}$-limit) and can describe the available NS data approximately equally well within about $2.5\lesssim\overline{\varepsilon}\lesssim4.5$,  the full $s^2$ simply increases monotonically with increasing $\overline{\varepsilon}$, as shown in panel (b) of FIG.\,\ref{fig_s2ab}.
Besides $k_1\approx0.006$ and $k_2\approx0.058$ in the revised scaling scheme,  we similarly have $k_1\approx0.037$ and $k_2\approx0.032$ for the original scaling scheme.
Within the effective data region in energy density, the parameterizations p1 and p2 behave very similarly and they are approximately equally accurate in describing the available NS data, as shown in FIG.\,\ref{fig_s2ab}.
It means that the currently available NS mass and radius data could not tell whether there exists a peaked $s^2$ profile within the energy density reachable in NS cores. So, 
can we extract a peaked speed of sound density profile near NS centers invariably from the currently available NS observational data? Unfortunately, our answer is no.

However, two related points warrant further discussion: (1) With the increasing availability of NS mass and radius measurements/observations, the uncertainties in $\Delta$ can be further reduced;
(2) Since our analysis is based on extracting the central energy densities using different NS instances, the resulting reduced energy density covers a relatively narrow range. Consequently, our main conclusion is most applicable to the SSS profile near the stellar core. The potential existence of a peaked structure farther from the center is beyond the scope of our current study and would require alternative methods or analyses.

\section{Summary and Outlook}\label{sec_conclusion}
In summary, within the IPAD-TOV approach based on the predicted scaling of NS compactness with its central pressure/energy density ratio that is also verified by $10^5$ meta-model EOSs, we have shown that NS central trace anomaly can be extracted reliably from its observational data directly insensitive of the EOS models. Using the available data on the mass, radius and/or compactness of several NSs from recent X-ray and gravitational wave observations, we extracted the central trace anomaly as a function of energy density. It stringently tests the existing EOS models and sets a clear guidance in a new direction for further understanding the nature and EOS of supradense matter in a model-insensitive manner. 

The central trace anomaly plays an important role in shaping the density or radius profile of the SSS $s^2$ in NSs.
Our analyses of the central trace anomaly 
indicate that a peaked shape of $s^2$ in NSs is an implication considering some well-founded theoretical limits but not practically inevitable, and it is not a direct consequence of the currently available NS observational data. 

\begin{figure}[tbh]
\centering
\includegraphics[width=8.cm]{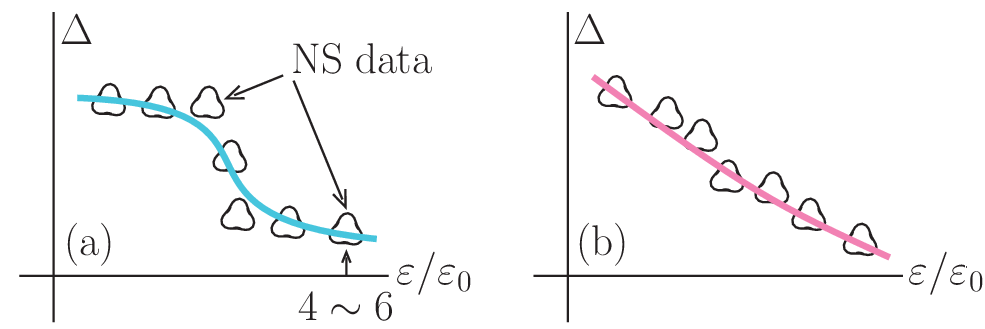}
\caption{(Color Online). The $\Delta$ in the left panel may generate a peaked $s^2$ while that in the right panel implies $\d\Delta/\d\overline{\varepsilon}\approx\rm{const.}<0$ and $s^2$ would thus increase monotonically with $\varepsilon/\varepsilon_0$. The solid line in each panel represents a probable shape of $\Delta$.
}\label{fig_DEL_s2peak}
\end{figure}

Looking forward, as more data of NS masses and radii or red-shifts become available, we may be able to finally determine whether a peaked $s^2$ profile can emerge in NSs. We emphasize that our scalings provide a useful tool to extract the $\Delta$ and $\varepsilon/\varepsilon_0$ directly from the observational data (masses, radii).
Sketched in FIG.\,\ref{fig_DEL_s2peak} are two typical shapes of the function $\Delta(\varepsilon/\varepsilon_0)$.
In the left panel, we expect the $s^2$ profile to have a peak somewhere since there is a quick decreasing of $\Delta$ and two approximate plateaus at low and high energy densities. While the $\Delta$ in the right panel indicates that $-\d\Delta/\d\overline{\varepsilon}$ is approximately a positive constant, both the derivative part $-\overline{\varepsilon}\d\Delta/\d\overline{\varepsilon}$ and the non-derivative term $1/3-\Delta=P/\varepsilon$ are thus expected to monotonically increase with $\varepsilon/\varepsilon_0$.

Regarding the $s^2$ as a function of energy density $\varepsilon$, different systems and/or calculations on $\Delta(\varepsilon)$ may set some effective constraints on the $s^2(\varepsilon)$ curve; e.g., the low-density CEFT\,\cite{Ess21} with $\Delta\lesssim3^{-1}$ at $\overline{\varepsilon}\lesssim1$ 
and high-density pQCD\,\cite{Bjorken83,Kurkela10,Gorda21PRL,Gorda23PRL} with $\Delta\gtrsim0$ at $\overline{\varepsilon}\gtrsim50$ set $s^2\approx\mathcal{O}(\phi)$ with $\phi\equiv P/\varepsilon$ and $s^2\approx3^{-1}$, respectively.
Since the maximum SSS in NSs is probably greater than 1/3\,\cite{Ecker22}, theoretically a peak may unavoidably emerge somewhere in the density profile of $s^2(\varepsilon)$ as $s^2$ approaches 1/3 at extremely high densities\,\cite{Bjorken83,Kurkela10,Gorda21PRL,Gorda23PRL}. Since NSs are still among the most mysterious objects in the Universe, 
no NS theory is completely reliable. Nevertheless, based on most theories and observations currently available\,\cite{Braun2022,Semp2025}, 
the position of the peak in $s^2(\varepsilon)$ profile may exceed the maximum densities reached in NSs. Hopefully, as more observational data on the mass, radius or red-shift becomes available, our approach based on NS compactness and mass scalings can help better constrain the trace anomaly $\Delta$ in a wide range of energy densities reachable in NSs, locate precisely the possible peak position of $s^2$ profile (within NS densities) relying on NS data alone without using other model dependent physics inputs\,\cite{Gorda23,ZhD25,Komo22}, and thus restrict the NS EOS to a much narrower band.

\section*{Acknowledgement}

We would like to thank Wen-Jie Xie, Nai-Bo Zhang and Zhen Zhang for helpful discussions. This work was supported in part by the U.S. Department of Energy, Office of Science, under Award Number DE-SC0013702, the CUSTIPEN (China-U.S. Theory Institute for Physics with Exotic Nuclei) under the US Department of Energy Grant No. DE-SC0009971.

\section*{DATA AVAILABILITY} The data that support the findings of this article will be openly available \cite{Datasets}.

\appendix

\section{Meta-model EOS of NS Matter}\label{app-meta}

We adopt a meta-model for generating randomly NS EOSs in a broad parameter space that can mimic diverse model predictions consistent will existing constraints from terrestrial experiments and astrophysical observations as well as general physics principles\,\cite{ZLX,ZLi-2,XL,ZLi-3}. 
It is based on a so-called minimum model of NSs consisting of neutrons, protons, electrons and muons (npe$\mu$ matter) at $\beta$-equilibrium. Its most basic input is the EOS of isospin-asymmetric nucleonic matter in the form of energy per nucleon $E(\rho,\delta)=E_0(\rho)+E_{\rm{sym}}(\rho)\delta^2$, here $\rho=\rho_{\rm{n}}+\rho_{\rm{p}}$ and $\delta\equiv(\rho_{\rm{n}}-\rho_{\rm{p}})/(\rho_{\rm{n}}+\rho_{\rm{p}})$ is the isospin asymmetry of neutron-rich system with neutron density $\rho_{\rm{n}}$ and proton density $\rho_{\rm{p}}$, respectively. The EOS of symmetric nuclear matter $E_0(\rho)$ and the symmetry energy $E_{\rm{sym}}(\rho)$ are parameterized respectively as $E_0(\rho)=B+2^{-1}K_0\chi^2+6^{-1}J_0\chi^3+24^{-1}I_0\chi^4$ and $E_{\rm{sym}}(\rho)=S+L\chi+2^{-1}K_{\rm{sym}}\chi^2+6^{-1}J_{\rm{sym}}\chi^3$ with $\chi\equiv(\rho-\rho_0)/3\rho_0$, $B\equiv E_0(\rho_0)$ and $S\equiv E_{\rm{sym}}(\rho_0)$, therefore defining the coefficients $K_0$, $J_0$, $\cdots$. 

The total energy density is $\varepsilon(\rho,\delta)=[E(\rho,\delta)+M_{\rm{N}}]\rho+\varepsilon_{\ell}(\rho,\delta)$ where $M_{\rm{N}}\approx939\,\rm{MeV}$ and $\varepsilon_{\ell}(\rho,\delta)$ is the energy density of leptons from an ideal Fermi gas model\,\cite{TOV39-2}. The pressure $P(\rho,\delta)$ is $P(\rho,\delta)=\rho^2\d[\varepsilon(\rho,\delta)/\rho]/\d\rho$. The density profile of isospin asymmetry $\delta(\rho)$ is obtained by solving the $\beta$-equilibrium condition $\mu_{\rm{n}}-\mu_{\rm{p}}=\mu_{\rm{e}}=\mu_\mu\approx4\delta E_{\rm{sym}}(\rho)$ and the charge neutrality requirement $\rho_{\rm{p}}=\rho_{\rm{e}}+\rho_\mu$. Here the chemical potential $\mu_i$ for a particle $i$ is calculated from the energy density via
$\mu_i=\partial\varepsilon(\rho,\delta)/\partial\rho_i.$ With the $\delta(\rho)$ calculated consistently using the inputs given above, both the pressure $P$ and energy density $\varepsilon(\rho)=\varepsilon(\rho, \delta(\rho))$ become barotropic, i.e., depend on the density $\rho$ only. The EOS in the form of $P(\varepsilon)$ can then be used in solving the TOV equations. 
The core-crust transition density $\rho_{\rm{cc}}$ is determined self-consistently by the thermodynamic method\,\cite{Lattimer:2006xb,XuJ}; in the inner crust with densities between $\rho_{\rm{cc}}$ and $\rho_{\rm{out}}\approx2.46\times10^{-4}\,\rm{fm}^{-3}$ corresponding to the neutron dripline we adopt the parametrized EOS $P=\alpha+\beta\varepsilon^{4/3}$\,\cite{Iida1997}; and for $\rho<\rho_{\rm{out}}$ we adopt the Baym-Pethick-Sutherland (BPS) and the Feynman-Metropolis-Teller (FMT) EOSs\,\cite{BPS71}.

In order for a broad verification of the scalings studied in this work, we select the saturation density $\rho_0$ to be $0.15\,\rm{fm}^{-3}\lesssim\rho_0\lesssim0.17\,\rm{fm}^{-3}$,  the binding energy in the range of $-17\,\rm{MeV}\lesssim B\lesssim-15\,\rm{MeV}$,  the incompressibility for symmetric matter in $210\,\rm{MeV}\lesssim K_0\lesssim250\,\rm{MeV}$; the skewness in $-400\,\rm{MeV}\lesssim J_0\lesssim0\,\rm{MeV}$ and the kurtosis in $400\,\rm{MeV}\lesssim I_0\lesssim1200\,\rm{MeV}$; for the symmetry energy we adopt $28\,\rm{MeV}\lesssim S\lesssim 36\,\rm{MeV}$, $30\,\rm{MeV}\lesssim L\lesssim90\,\rm{MeV}$, $-300\,\rm{MeV}\lesssim K_{\rm{sym}}\lesssim0\,\rm{MeV}$ as well as $200\,\rm{MeV}\lesssim J_{\rm{sym}}\lesssim 1000\,\rm{MeV}$.
These parameters generated randomly within the specified uncertainty ranges are consistent with terrestrial experimental and contemporary astrophysical constraints, see, e.g., Refs.\,\cite{ZLX,ZLi-2,XL,ZLi-3} for more discussions. A similar scheme on constructing NS EOS is based on the spectral representation\,\cite{Lindblom2010}, in which the linear combinations of certain selected basis functions are used. Our meta-model can fully mimic these EOSs. 
 
\begin{figure}[h!]
\centering
\includegraphics[width=8.5cm]{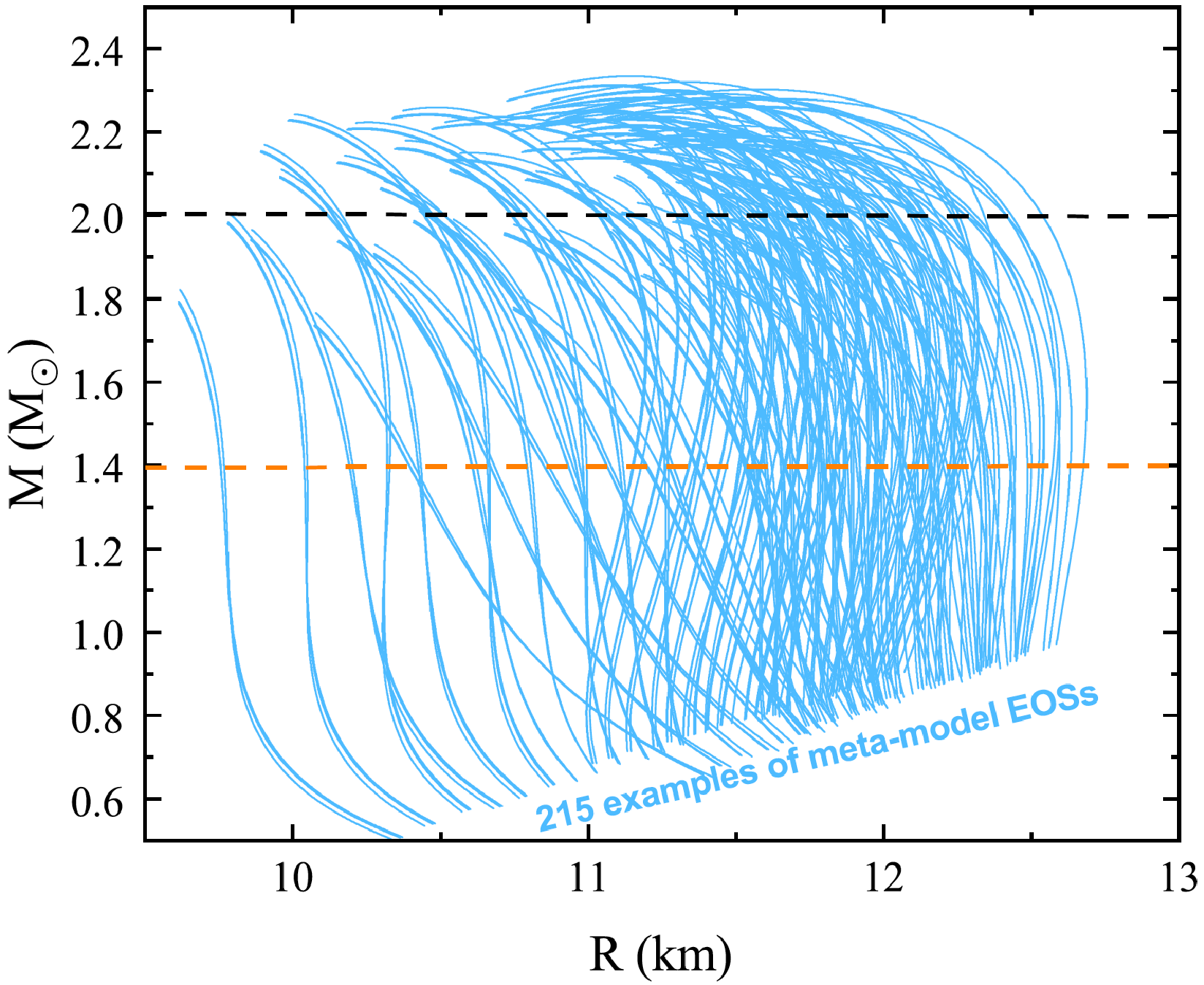}
\caption{(Color Online). 215 samples of the M-R curves generated by the meta-model NS EOSs.}\label{fig_MR-ensem}
\end{figure}

Shown in FIG.\,\ref{fig_MR-ensem} are 215 samples of the generated mass-radius curves using our meta-model EOSs\,\cite{Li:2024imk}. It is seen that the local derivative $\d M_{\rm{NS}}/\d R$ can switch broadly between negative and positive values. They have most of the features of the $M(R)$ generated using other meta-model (either with parameters randomly generated or specified), see, e.g.,  (1) the 40435 $M(R)$ curves from parameterizing the high-density EOS with piecewise polytropes\,\cite{Ferreira:2024hxc}; (2) samples  from using hybrid models coupling various hadronic EOSs through a first-order phase transition to quark matter EOSs characterized by different speeds of sound in studying the possible formation of twin stars\,\cite{Han:2020adu}; (3) samples with EOSs encapsulating quarkyonic matter in NSs\,\cite{Zhao20}; and (4) the 284 microscopic and/or phenomenological EOSs (see the next section). The $10^5$ EOSs we generated produce diverse $M(R)$ curves with features similar to those predicted by all of these EOSs.

\section{Derivation of the Correction ``$18\x/25$'' to Mass Scaling and its Improvement on NS Compactness Scaling}\label{app-correction}

Using our notations of Ref.\,\cite{CLZ23-a}, the dimensionless radius $\widehat{R}$ is obtained by truncating the equation $\widehat{P}\approx\x+b_2\widehat{R}^2\approx0$ where $b_2=-6^{-1}(1+3\x^2+4\x)$.
Therefore, $\widehat{R}=[6\x/(1+3\x^2+4\x)]^{1/2}=\sqrt{6\Pi_{\rm{c}}}\sim\Pi_{\rm{c}}^{1/2}$ with $\Pi_{\rm{c}}$ defined in Eq.\,(\ref{def_Pic}), and so $R\sim\varepsilon_{\rm{c}}^{-1/2}\widehat{R}\sim\varepsilon_{\rm{c}}^{-1/2}\Pi_{\rm{c}}^{1/2}$.
The dimensionless NS mass is given by $\widehat{M}_{\rm{NS}}\approx3^{-1}\widehat{R}^3+5^{-1}a_2\widehat{R}^5+\cdots$, here $a_2=b_2/s_{\rm{c}}^2$\,\cite{CLZ23-a} with $s_{\rm{c}}^2$ the central speed of sound squared.
Keeping only the leading-order term $3^{-1}\widehat{R}^3$ in $\widehat{M}_{\rm{NS}}$ gives the NS mass scaling $\widehat{M}_{\rm{NS}}\sim\Pi_{\rm{c}}^{3/2}$, so $M_{\rm{NS}}\sim\varepsilon_{\rm{c}}^{-1/2}\widehat{M}_{\rm{NS}}\sim\varepsilon_{\rm{c}}^{-1/2}\Pi_{\rm{c}}^{3/2}$\,\cite{CLZ23-a}.
The $s_{\rm{c}}^2$ for the TOV configuration could be derived by requiring $\d M_{\rm{NS}}/\d\varepsilon_{\rm{c}}=0$, this gives\,\cite{CLZ23-a}
\begin{equation}\label{def-sc2}
    s_{\rm{c}}^2=\x\left(1+\frac{1}{3}\frac{1+3\x^2+4\x}{1-3\x^2}\right).
\end{equation}

The term $5^{-1}a_2\widehat{R}^5$ in the expansion of NS mass plays the similar role as $b_2\widehat{R}^2$ in the expansion of pressure.
Including this term in $\widehat{M}_{\rm{NS}}$ enables us to write out
\begin{equation}\label{dk-1}
    \widehat{M}_{\rm{NS}}\approx\frac{1}{3}\widehat{R}^3\left(1+\frac{3}{5}a_2\widehat{R}^2\right)
    =\frac{1}{3}\widehat{R}^3\left(1-\frac{3}{5}\frac{\x}{s_{\rm{c}}^2}\right),
    \end{equation}
where the relations $\x+b_2\widehat{R}^2\approx0$ and $a_2=b_2/s_{\rm{c}}^2$ are used when obtaining the final expression.
The factor ``$1+3a_2\widehat{R}^2/5$'' is the average (reduced) energy density $\langle\widehat{\varepsilon}\rangle$ by including the $a_2$-term in $\widehat{\varepsilon}$ as  $\widehat{\varepsilon}(\widehat{r})\approx1+a_2\widehat{r}^2+\cdots$\,\cite{CLZ23-a,CL24-c}, namely $\widehat{M}_{\rm{NS}}\approx3^{-1}\widehat{R}^3\langle\widehat{\varepsilon}\rangle$ with
\begin{equation}
\langle\widehat{\varepsilon}\rangle=\left.\int_0^{\widehat{R}}\d\widehat{r}\widehat{r}^2\widehat{\varepsilon}(\widehat{r})\right/\int_0^{\widehat{R}}\d\widehat{r}\widehat{r}^2
    =1+\frac{3}{5}a_2\widehat{R}^2.
\end{equation}
Thus, the original mass scaling from the leading term of its expansion is equivalent to having used the central energy density $(\widehat{\varepsilon}=1)$ instead of the mean $\langle\widehat{\varepsilon}\rangle$. 

The $s_{\rm{c}}^2$ in Eq.\,(\ref{dk-1}) is no longer given by Eq.\,(\ref{def-sc2}), but should now include corrections due to including the $a_2$-term in the mass. Generally, we write it as\,\cite{CL24-c}:
\begin{align}\label{app-2}
    s_{\rm{c}}^2\approx&\x\left(1+\frac{1}{3}\frac{1+3\x^2+4\x}{1-3\x^2}\right)\left(1+\kappa_1\x\right)\notag\\
    \approx&\frac{4}{3}\x+\frac{4}{3}\left(1+\kappa_1\right)\x^2,
\end{align}
where $\kappa_1$ is a coefficient to be determined.
On the other hand, taking $\d M_{\rm{NS}}/\d\varepsilon_{\rm{c}}=0$ where $M_{\rm{NS}}\sim\varepsilon_{\rm{c}}^{-1/2}\widehat{M}_{\rm{NS}}$ with $\widehat{M}_{\rm{NS}}$ of Eq.\,(\ref{dk-1}) gives the expression for $s_{\rm{c}}^2$. We then expand the latter over $\x$ to order $\x^2$:
\begin{equation}\label{app-3}
    s_{\rm{c}}^2\approx\frac{4}{3}\x+\frac{1}{11}\left(\frac{38}{3}-2\kappa_1\right)\x^2.
\end{equation}
Matching the two expressions Eq.\,(\ref{app-2}) and Eq.\,(\ref{app-3}) at order $\x^2$ gives $\kappa_1=-3/25$.
Therefore,
\begin{equation}
    s_{\rm{c}}^2\approx\x\left(1+\frac{1}{3}\frac{1+3\x^2+4\x}{1-3\x^2}\right)\left(1-\frac{3}{25}\x\right),
\end{equation}
and so we determine that $\x\lesssim0.381$ via $s_{\rm{c}}^2\leq1$, which is close to and consistent with 0.374 obtained in Ref.\,\cite{CLZ23-a}; and similarly $\Delta\gtrsim-0.048$.
The magnitude of the correction ``$+\kappa_1\x$'' in $s_{\rm{c}}^2$ is smaller than 5\% while the corresponding correction on the upper bound of $\x$ is smaller than 2\%.
In addition, the NS mass now scales as
\begin{equation}
    M_{\rm{NS}}\sim
    \frac{1}{\sqrt{\varepsilon_{\rm{c}}}}\left(\frac{\x}{1+3\x^2+4\x}\right)^{3/2}\cdot\left(1+\frac{18}{25}\x\right),
\end{equation}
since $1-3\x/5s_{\rm{c}}^2\approx(11/20)[1+9(1+\kappa_1)\x/11]\sim1+18\x/25$.
The NS radius still scales as $R\sim\varepsilon_{\rm{c}}^{-1/2}\Pi_{\rm{c}}^{1/2}$ (keeping the $b_2$-term), and so 
$\xi=M_{\rm{NS}}/R\sim\Pi_{\rm{c}}(1+18\x/25)=\overline{\Pi}_{\rm{c}}$.

\begin{figure}[h!]
\centering
\includegraphics[width=8.5cm]{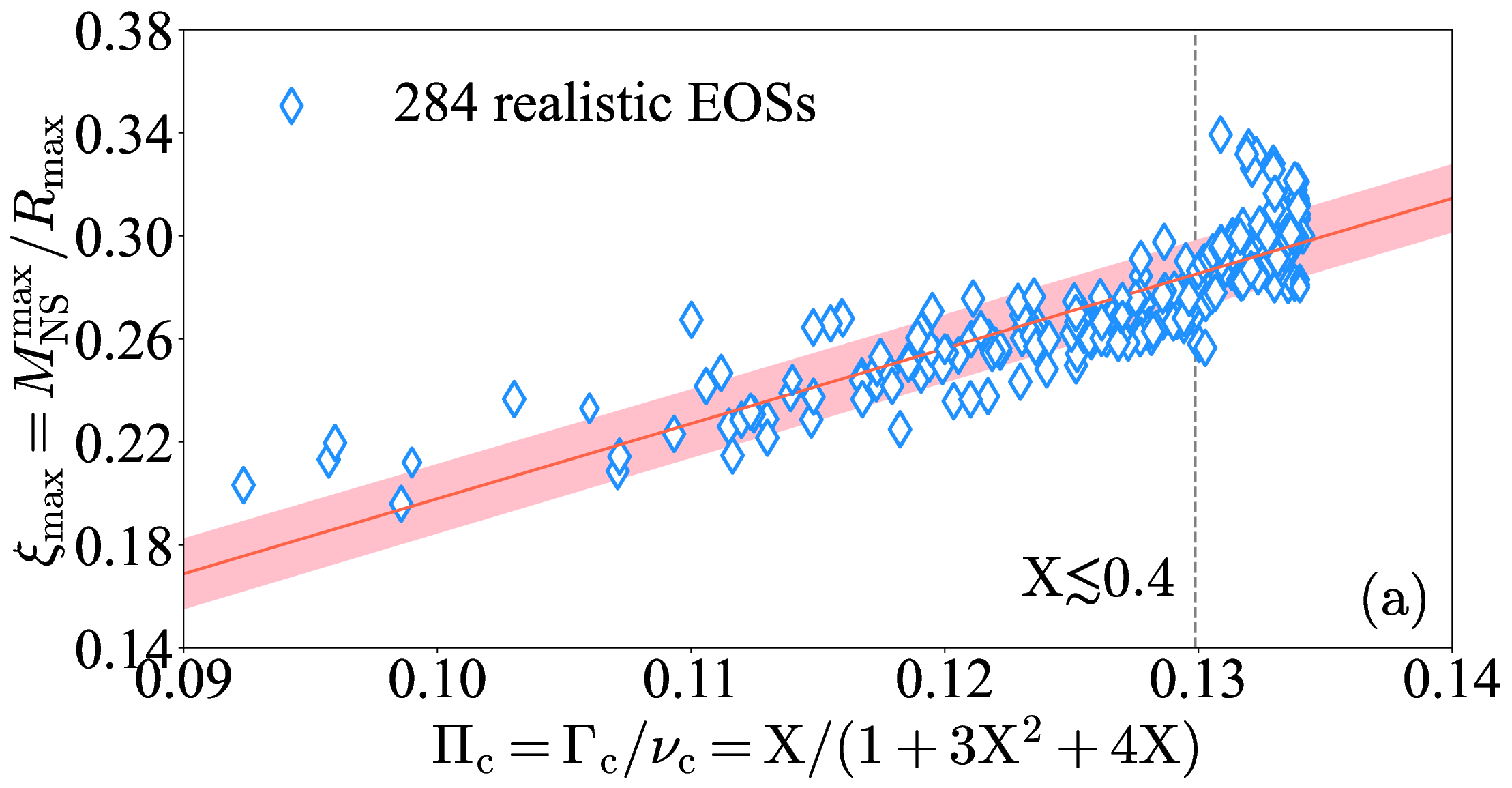}\quad
\includegraphics[width=8.5cm]{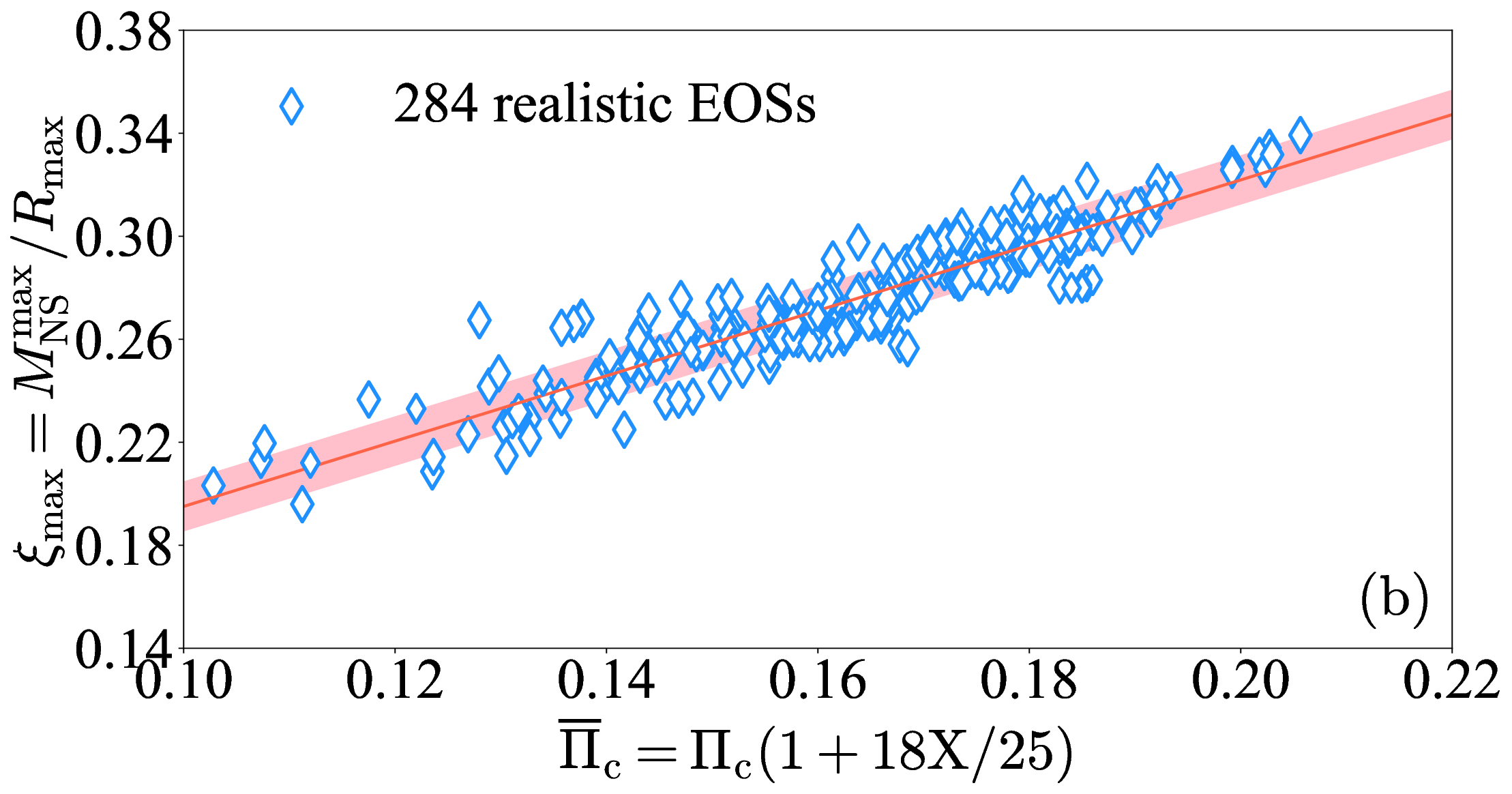}
\caption{(Color Online).  NS compactness scalings $M_{\rm{NS}}^{\max}/R_{\max}\sim\Pi_{\rm{c}}$ (upper panel) and $M_{\rm{NS}}^{\max}/R_{\max}\sim\overline{\Pi}_{\rm{c}}=\Pi_{\rm{c}}(1+18\x/25)$ (lower panel), using 284 realistic EOS samples, see text for details.
}\label{fig_rp_xi}
\end{figure}

Besides the improvement on the $\xi$-scaling by including the high-order correction $18\x/25$ given in the main text of FIG.\,\ref{fig_Pic},
here we study how this correction improves the compactness scaling for TOV NSs using 284 realistic EOSs. We note here by passing that the word "realistic" has been widely used in the literature to describe NS EOSs constructed based on predictions of some microscopic nuclear many-body theories and/or phenomenological energy density functionals.
Some of them have calculated the NS crust and core EOS consistently using the same interactions or energy density functionals. 
We emphasize that the $10^5$ meta-model EOS we used in this work are as realistic as these so-called "realistic" EOSs in terms of satisfying all available constraints from both terrestrial experiments and astrophysical observations. Nevertheless, following the naming tradition, we refer in the discussion here the following 284 EOSs as realistic EOSs to distinguish them from the meta-model EOSs. They can be classified generally as: (a) nucleonic models (microscopic/phenomenological); (b) hybrid EOS models with hyperons and/or $\Delta$ resonances; and (c) quark matter EOSs\,\cite{CLZ23-a}. 
In particular,  the following three types of EOS are included: (1) EOSs with a first-order phase transition, e.g., the APR EOS\,\cite{Akmal1998}, the EOSs based on chiral-mean-field (CMF) model (DS-CMF sereis)\,\cite{Dexheimer2021PRC} as well as the VQCD EOSs\,\cite{Jokela2021PRD}; (2) EOSs with a continuous hadron-quark crossover such as the AFL EOS series\,\cite{Alford2008} and the QHC EOSs\,\cite{Baym19}; and (3) EOSs whose SSS has multiple peaks using the quark-meson-coupling (QMC) model\,\cite{Stone2021MNRAS,Leong2024NPA} or the relativistic mean field (RMF) model with hyperons\,\cite{Stone2021MNRAS}, as well as the EOSs with multiple discontinuities by constructing sequential QCD phase transitions\,\cite{Alford2017PRL}.
See Refs.\,\cite{Typel2015,Ofeng24} for more details on these realistic EOSs.  

Shown in the upper panel of FIG.\,\ref{fig_rp_xi} are the correlations between $\xi_{\max}=M_{\rm{NS}}^{\max}/R_{\max}$ and $\Pi_{\rm{c}}$ (original mass scaling at the TOV configuration) using the 284 realistic EOSs. It is seen that for $\Pi_{\rm{c}}\lesssim0.13$ or equivalently $\x\lesssim0.4$ the scaling between $\xi_{\max}$ and $\Pi_{\rm{c}}$ is relatively good. Moreover, the nonlinearity appears only for even larger $\x\gtrsim0.4$ where $\x$'s may be in conflict with the upper bound on $\x$ due to causality.
In the lower panel of FIG.\,\ref{fig_rp_xi}, the correlation between the NS compactness and $\Pi_{\rm{c}}(1+18\x/25)$ (revised mass scaling at the TOV configuration) is shown using the same EOS samples. Clearly, including the correction $18\x/25$ is seen to significantly improve the overall correlation. More quantitatively, the correction $18\x/25$ for $\x\approx0.5$ is about 36\% of the leading term. It is sizable and could not be neglected. Indeed, the r-value of the fitting changes from 0.863 to about 0.942 after including the high-order correction $18\x/25$.

\begin{figure*}
\centering
\includegraphics[width=12.5cm]{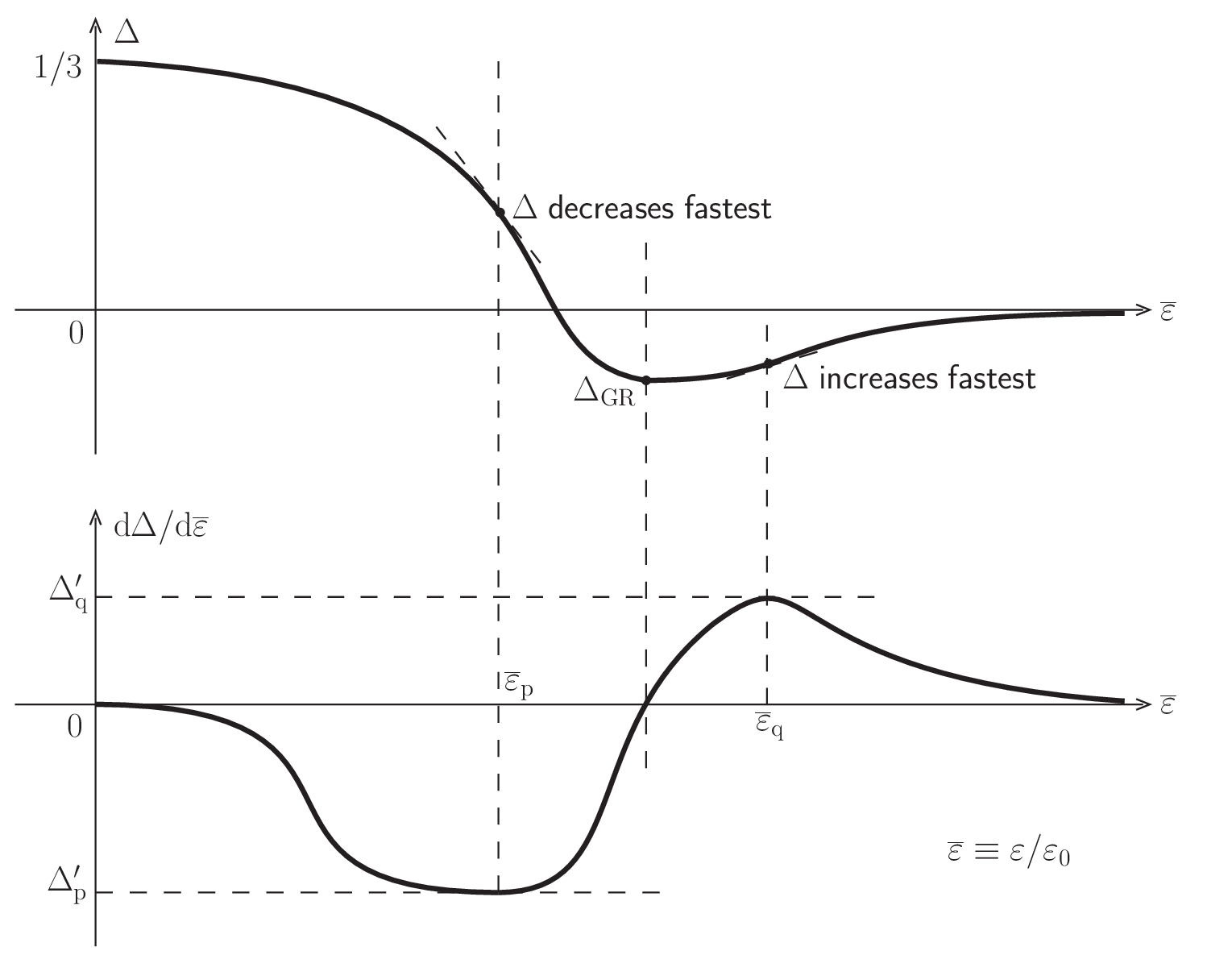}
\caption{Sketches of the trace anomaly $\Delta$ (upper panel) and its derivative $\d\Delta/\d\overline{\varepsilon}$ (lower panel) as functions of energy density.}\label{fig_s2phys}
\end{figure*}

\section{Understanding the Peaked/valleyed Behavior of $s^2$: Analytical Analyses}\label{app-s2}

\newcommand{\ep}{\overline{\varepsilon}}

The results of FIG.\,\ref{fig_s2ab} in the main text could be understood by considering the general feature of the derivative term $\Delta'\equiv\d\Delta/\d\overline{\varepsilon}$ of the dimensionless trace anomaly.
There is a point $\overline{\varepsilon}_{\rm{p}}$ in energy density where $\Delta$ decreases fastest; around this point we can expand $\Delta'$ as
\begin{equation}
\Delta'\approx\Delta_{\rm{p}}'+\frac{1}{2}\Delta_{\rm{p}}'''\left(\overline{\varepsilon}-\ep_{\rm{p}}\right)^2,~~
\Delta_{\rm{p}}'<0,~~\Delta_{\rm{p}}'''>0,
\end{equation}
since the first-order derivative of $\Delta'$, namely $\Delta''$ at such point is zero, see S.FIG.\,\ref{fig_s2phys}.
Since $\Delta_{\rm{p}}'''$ characterizes the curvature of $\Delta'$, it is small for a shallow-shaped $\Delta'$ but large for a sharp-shaped one. 
Correspondingly, the trace anomaly $\Delta$ itself around the point can be approximated by
\begin{equation}
\Delta\approx\Delta_{\rm{p}}'\left(\overline{\varepsilon}-\ep_{\rm{p}}\right)+\frac{1}{6}\Delta_{\rm{p}}'''\left(\overline{\varepsilon}-\ep_{\rm{p}}\right)^3+\Delta_{\rm{p}}.
\end{equation}
We obtain the $s^2$ using the formula (\ref{ss_decom}).
Calculating the derivative $\d s^2/\d\ep$ gives
\begin{equation}\label{for2}
{\d s^2}/{\d\ep}=-2\Delta_{\rm{p}}'+\left(
3\ep_{\rm{p}}\ep-2\ep^2-\ep_{\rm{p}}^2
\right)\Delta_{\rm{p}}'''.
\end{equation}

We discuss two cases:
(i) If the valley of the $\Delta'$ curve is shallow, i.e., $\Delta_{\rm{p}}'''$ is positively small, we can neglect the second term in (\ref{for2}),
\begin{equation}
{\d s^2}/{\d\ep}\approx-2\Delta_{\rm{p}}'>0.
\end{equation}
This means that even when there exists a peak in the derivative term $-\ep\d\Delta/\d\ep$ (equivalently a valley in $\d\Delta/\d\ep$),  the $s^2$ is still a monotonic increasing function of $\ep$.
This corresponds to the panel (b) of FIG.\,\ref{fig_s2ab} in the main text. (ii) On the other hand, if $|\Delta_{\rm{p}}'|$ is smaller than $\Delta_{\rm{p}}'''$,  i.e., the valley in $\Delta'$ is sharp, we can treat the first term in (\ref{ss_decom}) as a perturbation and solve the equation $\d s^2/\d\ep=0$ for its extreme point $\ep_{\rm{p}}^{\ast}$. The result is
\begin{align}\label{for3}
\ep_{\rm{p}}^{\ast}=&\frac{3\ep_{\rm{p}}\Delta_{\rm{p}}'''+\sqrt{\ep_{\rm{p}}^2\Delta_{\rm{p}}'''^2-16\Delta_{\rm{p}}'\Delta_{\rm{p}}'''}}{4\Delta_{\rm{p}}'''}\notag\\
\approx&\ep_{\rm{p}}\left(1-\frac{2}{\ep_{\rm{p}}^2}\frac{\Delta_{\rm{p}}'}{\Delta_{\rm{p}}'''}\right)>\ep_{\rm{p}},
\end{align}
the second approximation follows for small $\Delta_{\rm{p}}'$.
Moreover, we have
\begin{align}
s^2(\ep_{\rm{p}}^\ast)\approx&3^{-1}-\Delta_{\rm{p}}-\ep_{\rm{p}}\Delta_{\rm{p}}',~~
\mbox{as well as}\\
\left.\frac{\d^2s^2}{\d\ep^2}\right|_{\ep_{\rm{p}}^\ast}\approx&\Delta_{\rm{p}}'''\ep_{\rm{p}}\left(
\frac{8}{\ep_{\rm{p}}^2}
\frac{\Delta_{\rm{p}}'}{\Delta_{\rm{p}}'''}-1\right)<0.
\end{align}
The negativeness of the second-order derivative shows that it is a maximum point (peak) of $s^2$.
The correction in the bracket of (\ref{for3}) is positive, this means that the peak in $s^2$ occurs on the right side of the peak in its derivative term $-\ep\d\Delta/\d\ep$. See TAB.\,\ref{tab_X3} for an example, i.e, $5.2\approx\overline{\varepsilon}_{\rm{pk}}>\overline{\varepsilon}_{\rm{deriv,pk}}\approx4.1$;
this corresponds to the panel (a) of FIG.\,\ref{fig_s2ab} in the main text. We can also evaluate the decomposition terms $s_{\rm{deriv}}^2\equiv-\ep\d\Delta/\d\ep$ and $s_{\rm{non}\mbox{-}\rm{deriv}}^2\equiv 3^{-1}-\Delta$ of $s^2$ at $\ep_{\rm{p}}^\ast$,
namely $
s_{\rm{deriv}}^2(\ep_{\rm{p}}^\ast)\approx-\ep_{\rm{p}}\Delta_{\rm{p}}'$ and $s_{\rm{non}\mbox{-}\rm{deriv}}^2(\ep_{\rm{p}}^\ast)\approx
3^{-1}-\Delta_{\rm{p}}$.
Obviously, the derivative term is small positive.

The analysis for the point $\overline{\varepsilon}_{\rm{q}}$ in FIG.\,\ref{fig_s2phys} where $\Delta'$ is a maximum is totally parallel. In particular, there exists a valley in $s^2$,
\begin{equation}
\left.\frac{\d^2s^2}{\d\ep^2}\right|_{\ep_{\rm{q}}^\ast}\approx\Delta_{\rm{p}}'''\ep_{\rm{q}}\left(
\frac{8}{\ep_{\rm{q}}^2}
\frac{\Delta_{\rm{q}}'}{\Delta_{\rm{q}}'''}-1\right)>0,
\end{equation}
at
\begin{equation}
\ep_{\rm{q}}^\ast\approx
\ep_{\rm{q}}\left(1-\frac{2}{\ep_{\rm{q}}^2}\frac{\Delta_{\rm{q}}'}{\Delta_{\rm{q}}'''}\right)>\ep_{\rm{q}},
\end{equation}
since now $
\Delta_{\rm{q}}'>0$ and $\Delta_{\rm{q}}'''<0$.
This means that the valley in $s^2$ also appears after that in its derivative part, consistent with the numerical results given in TAB.\,\ref{tab_X3}, i.e., $13.7\approx\overline{\varepsilon}_{\rm{vl}}>\overline{\varepsilon}_{\rm{deriv,vl}}\approx13.2$.
The derivative part is definitely negative since $
s_{\rm{deriv}}^2(\ep_{\rm{q}}^\ast)\approx-\ep_{\rm{q}}\Delta_{\rm{q}}'<0$, 
see the dashed light-blue line in panel (a) of FIG.\,\ref{fig_s2ab} in the main text.
Considering $\d\Delta/\d\ep\to0$ for large $\ep$,  the $s^2$ finally approaches its asymptotic value determined by pQCD theories.

Finally, we study the relation between $\ep_{\rm{GR}}$, the position of a local minimum of $\Delta(\ep)$ which is assumed to be smaller than $\overline{\varepsilon}_{\rm{c}}$, and $\ep_{\rm{pk}}$, the peak position in $s^2$. According to the definition of $\Delta$, we have
\begin{align}
\frac{\d\Delta}{\d\ep}=-\frac{s^2-\phi}{\ep},~~
\frac{\d^2\Delta}{\d\ep^2}=\frac{1}{\ep^2}\left[2\left(s^2-\phi\right)-\ep\frac{\d s^2}{\d\ep}\right],
\end{align}
where $\phi=P/\varepsilon$.
At $\ep_{\rm{GR}}$, the first-order derivative is zero, i.e., $s_{\rm{GR}}^2=\phi_{\rm{GR}}$; while the second-order derivative is positive, this gives 
\begin{equation}
\frac{\d s^2}{\d\ep_{\rm{GR}}}<0.
\end{equation}
This means at $\ep_{\rm{GR}}$, the SSS is a decreasing function with respect to $\ep$. Equivalently, the peak position $\ep_{\rm{pk}}$ in $s^2$ should be smaller than $\ep_{\rm{GR}}$:
\begin{equation}\label{gk1}
\ep_{\rm{pk}}<\ep_{\rm{GR}}.
\end{equation}
Therefore, the position of a local minimum of $\Delta$ being smaller than $\overline{\varepsilon}_{\rm{c}}$ strongly implies the existence of a peak in $s^2$ within NS densities.
The inverse is generally not true: If there is a peak in $s^2$, the $\Delta$ within NS densities may or may not develop a local minimum, since $\ep_{\rm{pk}}<\overline{\varepsilon}_{\rm{c}}$ can not be used to infer $\ep_{\rm{GR}}<\overline{\varepsilon}_{\rm{c}}$, as clearly shown by  (\ref{gk1}).
These features are consistent with panels (a) and (a$'$) of FIG.\,\ref{fig_s2ab}, see the numerical values of TAB.\,\ref{tab_X3}.


\begin{references}
\fontdimen2\font=1.5pt
\bibitem{Walecka1974}J. Walecka, \href{https://www.sciencedirect.com/science/article/pii/0003491674902085}{Ann. Phys. \textbf{83}, 491 (1974).}

\bibitem{Chin1976}S. Chin, \href{https://www.sciencedirect.com/science/article/pii/0003491677900161}{Ann. Phys. \textbf{108}, 301 (1976).}

\bibitem{Freedman1977}B. Freedman and L. McLerran, \href{https://journals.aps.org/prd/abstract/10.1103/PhysRevD.16.1130}{Phys. Rev. D \textbf{16}, 1130 (1977);} \href{https://journals.aps.org/prd/abstract/10.1103/PhysRevD.16.1147}{1147 (1977);} \href{https://journals.aps.org/prd/abstract/10.1103/PhysRevD.16.1169}{1169 (1977).}

\bibitem{Akmal1998}A. Akmal, V. Pandharipande, and D. Ravenhall, \href{https://journals.aps.org/prc/abstract/10.1103/PhysRevC.58.1804}{Phys. Rev. C \textbf{58}, 1804 (1998).}

\bibitem{LP01}J. Lattimer and M. Prakash, \href{https://iopscience.iop.org/article/10.1086/319702}{Astrophys. J. \textbf{550}, 426 (2001).}


\bibitem{Alford2008}M. Alford {et al.}, \href{https://journals.aps.org/rmp/abstract/10.1103/RevModPhys.80.1455}{Rev. Mod. Phys. \textbf{80}, 1455 (2008).}

\bibitem{LCK08} B.A. Li, L.W. Chen, and C.M. Ko, \href{https://www.sciencedirect.com/science/article/pii/S0370157308001269}{Phys. Rep. \textbf{464}, 113 (2008).}


\bibitem{Bed15}P. Bedaque and A. Steiner, \href{https://journals.aps.org/prl/abstract/10.1103/PhysRevLett.114.031103}{Phys. Rev. Lett. \textbf{114}, 031103 (2015).}


\bibitem{Tews18} I. Tews {et al.}, \href{https://iopscience.iop.org/article/10.3847/1538-4357/aac267}{Astrophys. J. \textbf{860}, 149 (2018).}

\bibitem{McL19}L. McLerran and S. Reddy, \href{https://journals.aps.org/prl/abstract/10.1103/PhysRevLett.122.122701}{Phys. Rev. Lett. \textbf{122}, 122701 (2019).}

\bibitem{Baym19}G. Baym {et al.}, \href{https://iopscience.iop.org/article/10.3847/1538-4357/ab441e}{Astrophys. J. \textbf{885}, 42 (2019).}


\bibitem{Zhao20}T.Q. Zhao and J. Lattimer, \href{https://journals.aps.org/prd/abstract/10.1103/PhysRevD.102.023021}{Phys. Rev. D \textbf{102}, 023021 (2020).}

\bibitem{Tan22}H. Tan {et al.}, \href{https://journals.aps.org/prl/abstract/10.1103/PhysRevLett.128.161101}{Phys. Rev. Lett. \textbf{128}, 161101 (2022).}

\bibitem{Tan22-a}H. Tan {et al.}, \href{https://journals.aps.org/prd/abstract/10.1103/PhysRevD.105.023018}{Phys. Rev. D \textbf{105}, 023018 (2022).}

\bibitem{Alt22}S. Altiparmak, C. Ecker,  and L. Rezzolla, \href{https://iopscience.iop.org/article/10.3847/2041-8213/ac9b2a}{Astrophys. J. Lett. \textbf{939}, L34 (2022).}

\bibitem{Dri22}C. Drischler, S. Han, and S. Reddy, \href{https://journals.aps.org/prc/abstract/10.1103/PhysRevC.105.035808}{Phys. Rev. C \textbf{105}, 035808 (2022).}

\bibitem{Huang22}Y.J. Huang {et al.}, \href{https://journals.aps.org/prl/abstract/10.1103/PhysRevLett.129.181101}{Phys. Rev. Lett. \textbf{129}, 181101 (2022).}

\bibitem{Ecker22}C. Ecker and L. Rezzolla, \href{https://iopscience.iop.org/article/10.3847/2041-8213/ac8674/meta}{Astrophys. J. Lett. \textbf{939}, L35 (2022).}

\bibitem{Ecker22-a}C. Ecker and L.  Rezzolla, \href{https://academic.oup.com/mnras/article/519/2/2615/6957261?login=true}{Mon. Not. Roy. Astron. Soc. \textbf{519}, 2615 (2022).}

\bibitem{Musolino2024}C. Musolino, C. Ecker, and L.  Rezzolla,
\href{https://iopscience.iop.org/article/10.3847/1538-4357/ad1758}{Astrophys. J. 962 \textbf{61} (2024).}

\bibitem{Som23}R. Somasundaram, I. Tews,  and J. Margueron, \href{https://journals.aps.org/prc/abstract/10.1103/PhysRevC.107.025801}{Phys. Rev. C \textbf{107}, 025801 (2023).}


\bibitem{Pro23}C. Provid$\hat{\rm{e}}$ncia et al., \href{https://doi.org/10.1201/9781003306580}{Chapter 5 in the Book, 
Benhar, O., Lovato, A., Maselli, A., \& Pannarale, F. (Eds.). (2024). Nuclear Theory in the Age of Multimessenger Astronomy (1st ed.). CRC Press. }

\bibitem{Ann18}E. Annala {et al.}, \href{https://journals.aps.org/prl/abstract/10.1103/PhysRevLett.120.172703}{Phys. Rev. Lett. \textbf{120}, 172703 (2018).}

\bibitem{Ann23}E. Annala et al., \href{https://www.nature.com/articles/s41467-023-44051-y}{Nat. Comm. \textbf{14}, 8451 (2023).}


\bibitem{ZLi23}N.B. Zhang and B.A.  Li, \href{https://link.springer.com/article/10.1140/epja/s10050-023-01010-x}{Eur. Phys. J. A  \textbf{59}, 86 (2023).}

\bibitem{Cao23}Z. Cao and L.W. Chen, \href{https://arxiv.org/pdf/2308.16783.pdf}{arXiv:2308.16783 (2023).}

\bibitem{Mro23}D. Mroczek et al.,  \href{https://arxiv.org/pdf/2309.02345.pdf}{Phys. Rev. D \textbf{110}, 123009 (2024). }


\bibitem{Ess21}R.  Essick et al., \href{https://journals.aps.org/prl/abstract/10.1103/PhysRevLett.127.192701}{Phys. Rev. Lett. \textbf{127}, 192701 (2021).}

\bibitem{Brandes23}L. Brandes, W. Weise, and N. Kaiser,
\href{https://journals.aps.org/prd/abstract/10.1103/PhysRevD.107.014011}{Phys. Rev. D \textbf{107}, 014011 (2023).}

\bibitem{Brandes23-a}L. Brandes, W. Weise, and N. Kaiser, \href{https://journals.aps.org/prd/abstract/10.1103/PhysRevD.108.094014}{Phys. Rev. D \textbf{108}, 094014 (2023).}

\bibitem{Tak23}J. Takatsy et al., \href{https://journals.aps.org/prd/abstract/10.1103/PhysRevD.108.043002}{Phys. Rev. D \textbf{108}, 043002 (2023).}

\bibitem{Pang23}P. Pang et al., \href{https://journals.aps.org/prc/abstract/10.1103/PhysRevC.109.025807}{Phys. Rev. C \textbf{109}, 025807 (2024).}

\bibitem{Fan23}Y.Z. Fan et al., \href{https://journals.aps.org/prd/abstract/10.1103/PhysRevD.109.043052}{Phys. Rev. D \textbf{109}, 043502 (2024).}



\bibitem{Olii23}D. Oliinychenko et al., \href{https://journals.aps.org/prc/pdf/10.1103/PhysRevC.108.034908}{Phys. Rev. C \textbf{108}, 034908 (2023).}

\bibitem{Wat16}A. Watts et al., \href{https://journals.aps.org/rmp/abstract/10.1103/RevModPhys.88.021001}{Rev. Mod. Phys. \textbf{88}, 021001 (2016).}

\bibitem{Oertel2017} M. Oertel {et al.}, \href{https://journals.aps.org/rmp/abstract/10.1103/RevModPhys.89.015007}{Rev. Mod. Phys.  \textbf{89}, 015007 (2017).}


\bibitem{Baym18}G. Baym {et al.}, \href{https://iopscience.iop.org/article/10.1088/1361-6633/aaae14}{Rep. Prog. Phys. \textbf{81}, 056902 (2018).}

\bibitem{Isa18}I. Vida$\widetilde{\rm{n}}$a, \href{https://royalsocietypublishing.org/doi/10.1098/rspa.2018.0145}{Proc. Roy. Soc. Lond. A \textbf{474}, 0145 (2018).}

\bibitem{LCXZ21}B.A. Li et al., \href{https://www.mdpi.com/2218-1997/7/6/182}{Universe \textbf{7}, 182 (2021).}



\bibitem{Dri21}C. Drischler, J. Holt, and C. Wellenhofer, \href{https://www.annualreviews.org/doi/10.1146/annurev-nucl-102419-041903}{Annu. Rev. Nucl. Part. Sci. \textbf{71}, 403 (2021).}

\bibitem{Lovato22}A. Lovato {et al.},  \href{https://arxiv.org/abs/2211.02224}{arXiv:2211.02224 (2022).}

\bibitem{Soren2023}A. Sorensen {et al.}, \href{https://doi.org/10.1016/j.ppnp.2023.104080}{Prog. Part. Nucl. Phys.  \textbf{134}, 104080 (2024).}

\bibitem{Fuji22}Y. Fujimoto {et al.}, \href{https://journals.aps.org/prl/abstract/10.1103/PhysRevLett.129.252702}{Phys. Rev. Lett. \textbf{129}, 252702 (2022).}

\bibitem{Riley19}T. Riley et al., \href{https://iopscience.iop.org/article/10.3847/2041-8213/ab481c}{Astrophys. J. Lett. \textbf{887}, L21 (2019).}

\bibitem{Miller19}M. Miller {et al.}, \href{https://iopscience.iop.org/article/10.3847/2041-8213/ab50c5}{Astrophys. J. Lett. \textbf{887}, L24 (2019).}


\bibitem{Fon21} E. Fonseca {et al.},  \href{https://iopscience.iop.org/article/10.3847/2041-8213/ac03b8}{Astrophys. J. Lett. \textbf{915}, L12 (2021).}

\bibitem{Riley21}T. Riley {et al.}, \href{https://iopscience.iop.org/article/10.3847/2041-8213/ac0a81}{Astrophys. J. Lett. \textbf{918}, L27 (2021).}

\bibitem{Miller21}M. Miller {et al.}, \href{https://iopscience.iop.org/article/10.3847/2041-8213/ac089b}{Astrophys. J. Lett. \textbf{918}, L28 (2021).}

\bibitem{Salmi22}T. Salmi et al., \href{https://iopscience.iop.org/article/10.3847/1538-4357/ac983d/pdf}{Astrophys. J. \textbf{941}, 150 (2022).}

\bibitem{Choud24}D. Choudhury et al., \href{https://iopscience.iop.org/article/10.3847/2041-8213/ad5a6f}{Astrophys. J.  Lett. \textbf{971}, L20 (2024).}

\bibitem{Reardon24}D. Reardon et al., \href{https://iopscience.iop.org/article/10.3847/2041-8213/ad614a}{Astrophys. J. Lett. \textbf{971}, L18 (2024).}


\bibitem{Zhou23}X. Zhou et al., \href{https://www.nature.com/articles/s41567-023-02034-2}{Nat. Phys. \textbf{1477}, 69 (2023).}

\bibitem{Abbott2017}B. Abbott {et al.}, \href{https://journals.aps.org/prl/abstract/10.1103/PhysRevLett.119.161101}{Phys. Rev. Lett. \textbf{119}, 161101 (2017).}

\bibitem{Abbott2018}B. Abbott {et al.}, \href{https://journals.aps.org/prl/abstract/10.1103/PhysRevLett.121.161101}{Phys. Rev. Lett. \textbf{121}, 161101 (2018).}



\bibitem{Abbott2020-a}B. Abbott et al., \href{https://iopscience.iop.org/article/10.3847/2041-8213/ab75f5}{Astrophys. J. Lett. \textbf{892}, L3 (2020).}


\bibitem{Sul24}A. Sullivan and R. Romani, \href{https://iopscience.iop.org/article/10.3847/1538-4357/ad4d85}{AstroPhys. J. \textbf{974}, 315 (2024).}

\bibitem{CL25}B.J. Cai and B.A. Li, \href{https://link.springer.com/article/10.1140/epja/s10050-025-01507-7}{Eur. Phys. J. A \textbf{61}, 55 (2025).}
\bibitem{TOV39-1}R. Tolman, \href{https://journals.aps.org/pr/abstract/10.1103/PhysRev.55.364}{Phys. Rev. \textbf{55}, 364 (1939).}
\bibitem{TOV39-2}
J. Oppenheimer and G. Volkoff, \href{https://journals.aps.org/pr/abstract/10.1103/PhysRev.55.374}{Phys. Rev. \textbf{55}, 374 (1939).}

\bibitem{CLZ23-a}B.J. Cai, B.A. Li, and Z. Zhang, \href{https://iopscience.iop.org/article/10.3847/1538-4357/acdef0}{Astrophys. J. \textbf{952}, 147 (2023).}
\bibitem{CLZ23-b}B.J. Cai, B.A. Li, and Z. Zhang, \href{https://doi.org/10.1103/PhysRevD.108.103041}{Phys. Rev. D \textbf{108}, 103041 (2023).}


\bibitem{CL24}B.J. Cai and B.A. Li, \href{https://doi.org/10.1103/PhysRevD.109.083015}{Phys. Rev. D \textbf{109}, 083015 (2024).}

\bibitem{Lat24-talk}J. Lattimer,  slide 15, talk given at the \href{https://indico.mitp.uni-mainz.de/event/380/contributions/4746/attachments/3474/4433/Bormio24Lattimer.pdf}{60th International Winter Meeting in Nuclear Physics, Jan. 22-26, 2024, Bormio, Italy.}

\bibitem{Ofeng20}D. Ofengeim, \href{https://journals.aps.org/prd/abstract/10.1103/PhysRevD.101.103029}{Phys. Rev. D \textbf{101}, 103029 (2020).}

\bibitem{Ofeng24}D. Ofengeim, P. Shternin,  and T. Piran, \href{https://journals.aps.org/prd/abstract/10.1103/PhysRevD.110.103046}{Phys.Rev. D \textbf{110}, 103046 (2024).}

\bibitem{SL24}B.Y. Sun and J. Lattimer, \href{https://iopscience.iop.org/article/10.3847/1538-4357/adc25d}{Astrophys. J. \textbf{984}, 30 (2025).}

\bibitem{ZLX}N.B. Zhang, B.A.  Li, and J. Xu, \href{https://iopscience.iop.org/article/10.3847/1538-4357/aac027}{Astrophys. J. \textbf{859}, 90 (2018)}


\bibitem{ZLi-2}N.B. Zhang and B.A.  Li,  \href{https://iopscience.iop.org/article/10.3847/1538-4357/ab24cb}{Astrophys. J. \textbf{879}, 99 (2019).}
\bibitem{XL}W.J. Xie and B.A. Li, \href{https://iopscience.iop.org/article/10.3847/1538-4357/aba271}{Astrophys. J. \textbf{899}, 4 (2020).}
\bibitem{ZLi-3}N.B. Zhang and B.A.  Li,   \href{https://iopscience.iop.org/article/10.3847/1538-4357/ac1e8c/meta}{Astrophys. J. \textbf{921}, 111 (2021).}

\bibitem{Lindblom2010}L. Lindblom, \href{https://journals.aps.org/prd/abstract/10.1103/PhysRevD.82.103011}{Phys. Rev. D \textbf{82}, 103011 (2010).}


\bibitem{Ann22}E. Annala et al., \href{https://journals.aps.org/prx/abstract/10.1103/PhysRevX.12.011058}{Phys. Rev. X \textbf{12}, 011058 (2022).}

\bibitem{Rich23}J. Richter and B.A. Li, \href{https://journals.aps.org/prc/abstract/10.1103/PhysRevC.108.055803}{Phys. Rev. C \textbf{108}, 055803 (2023).}

\bibitem{Abbott2021}B. Abbott et al., \href{https://doi.org/10.1103/PhysRevX.11.021053}{Phys. Rev. X \textbf{11}, 021053 (2021).}



\bibitem{Gorda23}T. Gorda, O. Komoltsev, and A. Kurkela, \href{https://iopscience.iop.org/article/10.3847/1538-4357/acce3a}{Astrophys. J. \textbf{950}, 107 (2023).}



\bibitem{Mam2021}M. Al-Mamun {et al.}, \href{https://journals.aps.org/prl/abstract/10.1103/PhysRevLett.126.061101}{Phys. Rev. Lett. \textbf{126},  061101 (2021).}



\bibitem{Bjorken83}J. Bjorken, \href{https://journals.aps.org/prd/abstract/10.1103/PhysRevD.27.140}{Phys. Rev. D \textbf{27}, 140 (1983).}

\bibitem{Kurkela10}A. Kurkela, P. Romatschke, and A. Vuorinen, \href{https://journals.aps.org/prd/abstract/10.1103/PhysRevD.81.105021}{Phys. Rev. D \textbf{81}, 105021 (2010).}

\bibitem{Gorda21PRL}T. Gorda et al., \href{https://journals.aps.org/prl/abstract/10.1103/PhysRevLett.127.162003}{Phys. Rev. Lett. \textbf{127}, 162003 (2021).}

\bibitem{Gorda23PRL}T. Gorda et al., \href{https://journals.aps.org/prl/abstract/10.1103/PhysRevLett.131.181902}{Phys. Rev. Lett. \textbf{131}, 181902 (2023).}

\bibitem{Braun2022}J. Braun and B. Schallmo, \href{https://journals.aps.org/prd/abstract/10.1103/PhysRevD.106.076010}{Phys. Rev. D \textbf{106}, 076010 (2022).}

\bibitem{Semp2025}A. Semposki et al., \href{https://journals.aps.org/prc/abstract/10.1103/PhysRevC.111.035804}{Phys. Rev. C \textbf{111}, 035804 (2025).}

\bibitem{ZhD25}D. Zhou, \href{https://journals.aps.org/prc/abstract/10.1103/PhysRevC.111.015810}{Phys. Rev. C \textbf{111}, 015810 (2025).}

\bibitem{Komo22}O. Komoltsev and A. Kurkela, \href{https://journals.aps.org/prl/abstract/10.1103/PhysRevLett.128.202701}{Phys. Rev. Lett. \textbf{128}, 202701 (2022).}

\bibitem{Dexheimer2021PRC}V. Dexheimer et al., \href{https://journals.aps.org/prc/abstract/10.1103/PhysRevC.103.025808}{Phys. Rev. C \textbf{103}, 0258 (2021).}

\bibitem{Jokela2021PRD}N. Jokela et al., \href{https://journals.aps.org/prd/abstract/10.1103/PhysRevD.103.086004}{Phys. Rev. C \textbf{103}, 086004 (2021).}

\bibitem{Stone2021MNRAS}J. Stone et al., \href{https://academic.oup.com/mnras/article/502/3/3476/6061395}{Mon. Not. Roy. Astron. Soc. \textbf{502}, 3476 (2021).}

\bibitem{Leong2024NPA}J. Leong, 
A. Thomas, and P. Guichon, \href{https://www.sciencedirect.com/science/article/pii/S0375947424001106}{Nucl. Phys. \textbf{A1050}, 122928 (2024).}

\bibitem{Alford2017PRL}M. Alford and A. Sedrakian, \href{https://journals.aps.org/prl/abstract/10.1103/PhysRevLett.119.161104}{Phys. Rev. Lett. \textbf{119}, 161104 (2017).}

\bibitem{Typel2015}
S. Typel, M. Oertel, and T. Klahn, \href{https://link.springer.com/article/10.1134/S1063779615040061}{ Physics of Particles and Nuclei \textbf{46}, 633 (2015).}

\bibitem{CL24-c}B.J. Cai and B.A. Li, \href{https://www.frontiersin.org/journals/astronomy-and-space-sciences/articles/10.3389/fspas.2024.1502888/full}{Front.  Astron.  Space Sci., Vol.  \textbf{11} (2024).}


\bibitem{Lattimer:2006xb}
J. Lattimer and M. Prakash,  \href {http://dx.doi.org/10.1016/j.physrep.2007.02.003} {Phys. Rep.  {\bf442}, 109 (2007).}
\bibitem{XuJ}J. Xu et al.,   \href{https://iopscience.iop.org/article/10.1088/0004-637X/697/2/1549}{Astrophys. J. \textbf{697}, 1549 (2009).}

\bibitem{Iida1997} K. Iida and K. Sato, \href{https://iopscience.iop.org/article/10.1086/303685}{Astrophys. J. \textbf{477},  294 (1997).}


\bibitem{BPS71} G. Baym, C. Pethick, and P. Sutherland, \href{https://articles.adsabs.harvard.edu/full/1971ApJ...170..299B}{Astrophys. J. \textbf{170}, 299 (1971).} 

\bibitem{Li:2024imk}
B.A. Li et al.,
\href{https://journals.aps.org/prd/abstract/10.1103/PhysRevD.110.103040}
{Phys. Rev. D \textbf{110}, 103040 (2024).}

\bibitem{Ferreira:2024hxc}
M. Ferreira and C. Provid\^encia,
\href{https://journals.aps.org/prd/abstract/10.1103/PhysRevD.110.063018}
{Phys. Rev. D \textbf{110},  063018 (2024).}

\bibitem{Han:2020adu}
S. Han and M. Prakash,
\href{https://iopscience.iop.org/article/10.3847/1538-4357/aba3c7}{
Astrophys. J. \textbf{899}, 164 (2020).}

\bibitem{Datasets}Bao-Jun Cai and Bao-An Li,\\
Datasets used in all figures will be available at
\url{https://doi.org/10.7910/DVN/AGDQHC},\\ Harvard Dataverse, V1

\end{references}
\end{document}